\documentclass[aps,twocolumn,superscriptaddress,preprintnumbers,floats]{revtex4}
\usepackage[colorlinks, citecolor=blue,anchorcolor=red,menucolor=red, linkcolor=red,filecolor=red,urlcolor=blue,frenchlinks=red]{hyperref}
\usepackage{amsfonts}
\usepackage{amsmath}
\usepackage{amssymb}
\usepackage{CJKutf8}
\usepackage{color}
\usepackage{comment}
\usepackage{epsfig}
\usepackage{epstopdf}
\usepackage{float}
\usepackage{graphicx,booktabs}
\usepackage{indentfirst}
\usepackage{longtable,lscape}
\usepackage{mathrsfs}
\usepackage{mathtools}
\usepackage{morefloats}
\usepackage{pifont}
\usepackage{txfonts}
\usepackage{multirow}

\begin{document}

\title{Higher excited charmed and charmed-strange mesons in an unquenched quark model}

\author{Ru-Hui Ni}
\email{niruhui@ucas.ac.cn}
\affiliation{School of Physical Sciences, University of Chinese Academy of Sciences, Beijing 100049, China}
\author{Jia-Jun Wu}
\email{wujiajun@ucas.ac.cn}
\affiliation{School of Physical Sciences, University of Chinese Academy of Sciences, Beijing 100049, China}
\affiliation{Southern Center for Nuclear-Science Theory (SCNT), Institute of Modern Physics, Chinese Academy of Sciences, Huizhou 516000, China}
\author{Xian-Hui Zhong}
\email{zhongxh@hunnu.edu.cn}
\affiliation{Department of Physics, Hunan Normal University, and Key Laboratory of Low-Dimensional Quantum Structures and Quantum Control of Ministry of Education, Changsha 410081, China}
\affiliation{Synergetic Innovation Center for Quantum Effects and Applications (SICQEA), Hunan Normal University, Changsha 410081, China}

\begin{abstract}
In this paper, as a continuation of our previous work on the low-lying heavy-light meson spectra, we systematically study the mass spectra and Okubo-Zweig-Iizuka-allowed strong decays of the higher $3S$-, $2P$-, $2D$-, and $1F$-wave charmed and charmed-strange mesons within a unified unquenched quark model.
It is found that coupled-channel effects induce sizable state dependent mass shifts, which modify the relative level spacings and lower the predicted masses compared with conventional quenched calculations.
The newly observed $D_{s1}(2933)^+$ reported by the LHCb collaboration could be identified as the low mass axial vector state $D_s(2P_1)$ via the $2^1P_1-2^3P_1$ mixing.
For the broad structure $D_{sJ}(3040)^+$ observed earlier by the \emph{BABAR} collaboration, the $D_s(3^1S_0)$ assignment seems to be favored over the high mass mixed state $D_s(2P_1^\prime)$.
Meanwhile, the $D(3000)^0$ signals reported by LHCb cannot be consistently explained by any of the considered $3S$, $2P$, $2D$, or $1F$ assignments.
Our predicted masses and decay properties of the missing higher $D$ and $D_s$ mesons may provide useful information for future experimental searches.
\end{abstract}


\maketitle

\section{Introduction}\label{sec:intro}

Since 2006, great progress has been achieved in searching for the excited charmed and charmed-strange mesons in experiments.
For the charmed meson sector, the \emph{BABAR} collaboration~\cite{BaBar:2010zpy} first observed several resonances, which are known as $D_0(2550)$, $D^*_1(2600)$, $D_2(2740)$, and $D^*_3(2750)$, in the $D^{(*)}\pi$ invariant mass spectra.
In an amplitude analysis of $B^-\to D^{(*)+}\pi^-\pi^-$ decays, the spin-parity numbers of $D_0(2550)$, $D^*_1(2600)$, $D_2(2740)$, and $D^*_3(2750)$ were determined by the LHCb collaboration~\cite{LHCb:2016lxy,LHCb:2019juy}.
Furthermore, the LHCb collaboration also observed new resonance structures, denoted as $D(3000)^0$ by the Particle Data Group (PDG)~\cite{ParticleDataGroup:2024cfk}, in the $D^{(*)}\pi$ invariant mass spectra around the mass region of $3.0$ GeV~\cite{LHCb:2013jjb,LHCb:2016lxy}.
In theory, the $D_0(2550)$ and $D^*_1(2600)$ can be well explained as the $2S$-wave $D$-meson excitations, while the $D_2(2740)$ and $D^*_3(2750)$ are consistent with the assignments of the $1D$-wave excitations~\cite{Sun:2010pg,Wang:2010ydc,Chen:2011rr,Li:2010vx,Badalian:2011tb,Lu:2014zua,Song:2015fha,Xiao:2014ura,Li:2017zng,Wang:2016krl,Yu:2016mez,Gupta:2018zlg}.
However, the nature of the $D(3000)^0$ structures around $3.0~\mathrm{GeV}$ is still controversial due to limited experimental information.
Different assignments, such as $3S$, $2P$, and $1F$, have been proposed in various theoretical studies~\cite{Wang:2013tka,Sun:2013qca,Yu:2014dda,Lu:2014zua,Xiao:2014ura,Song:2015fha,Godfrey:2015dva,Li:2017zng,Gandhi:2019lta,Wang:2016krl,Yu:2016mez,Gupta:2018zlg}.

For the charmed-strange meson sector, in 2006, the \emph{BABAR} collaboration reported a new narrow structure denoted as $D_{sJ}(2860)$ and a broad enhancement around $2.7~\mathrm{GeV}$ in the $DK$ invariant mass spectrum~\cite{BaBar:2006gme}.
Subsequently, the Belle collaboration established the structure around $2.7$ GeV, known as $D_{s1}(2700)$~\cite{ParticleDataGroup:2024cfk}, in $B^+\to\bar D^0 D^0 K^+$ decays and determined its spin-parity numbers to be $J^P=1^-$~\cite{Belle:2007hht}.
In a subsequent Dalitz plot analysis of $B_s^0\to\bar D^0 K^-\pi^+$ decays, the LHCb collaboration resolved the structure near $2.86~\mathrm{GeV}$ into two highly overlapping resonances with $J^P=1^-$ and $3^-$, $D_{s1}^*(2860)$ and $D_{s3}^*(2860)$~\cite{LHCb:2014ott,LHCb:2014ioa}. The $D_{s3}^*(2860)$ was also observed in the $D^{*+}K_S^0$ and $D^{*0}K^+$ final states~\cite{LHCb:2016mbk}.
The $D_{s1}(2700)$ may be assigned as the $D_s(2^3S_1)$ state or a mixed state with a sizeable component of the $1^3D_1$ state~\cite{Close:2006gr,Zhang:2006yj,Chen:2009zt,Zhong:2009sk,Song:2015nia,Li:2017sww}.
The $D_{s1}^*(2860)$ and $D_{s3}^*(2860)$ can be assigned as the $1D$ charmed-strange states, while the vector state may have sizeable mixing with the $2^3S_1$ state~\cite{Zhong:2009sk,Song:2015nia,Segovia:2015dia,Godfrey:2015dva}.
In 2009, in the high mass region, a broad $D_{sJ}(3040)^+$ structure was observed by the \emph{BABAR} collaboration in the $D^*K$ invariant mass spectrum~\cite{BaBar:2009rro}.
This structure may be attributed to the $2P_1$ excitations in the $D_s$-meson family as discussed in the literature~\cite{Sun:2009tg,Colangelo:2010te,Chen:2009zt,Zhong:2009sk,Xiao:2014ura,Song:2015nia,Badalian:2011tb,Li:2017zng,Jiang:2024ftx}.
However, its nature remains unclear due to a lack of experimental information.
In 2021, a new resonance $D_{s0}(2590)^+$ with $J^P=0^-$ was observed by the LHCb collaboration~\cite{LHCb:2020gnv}, which may be assigned
as the radial excitation $D_s(2^1S_0)$~\cite{Hao:2022vwt,Yang:2023tvc}.
Very recently, the LHCb collaboration observed a new excited charmed-strange meson, $D_{s1}(2933)^+$, in the $B^0\to D^+D^-K^+\pi^-$ decay~\cite{LHCb:2026sup}.
Its spin-parity was determined to be $J^P=1^+$, with a measured mass $M_{\rm exp}=2933^{+6+4}_{-5-3}~\mathrm{MeV}$ and width $\Gamma_{\rm exp}=72^{+18+7}_{-12-10}~\mathrm{MeV}$~\cite{LHCb:2026sup}.
Compared with the quark model predictions, the $D_{s1}(2933)^+$ is a good candidate for the higher excitation $D_s(2P)$ with $J^P=1^+$.

The discovery of $D_{s1}(2933)^+$ may be a milestone in establishing the higher excitations, such as the $3S$, $2P$, $2D$, and $1F$ states, in the $D$- and $D_s$-meson families.
To understand the nature of the higher excitations in the $D$- and $D_s$-meson families, many theoretical studies have been carried out.
For example, the mass spectra have been calculated within various quark models~\cite{Badalian:2011tb,Godfrey:2015dva,Song:2015nia,Song:2015fha,Ni:2021pce}, and their Okubo-Zweig-Iizuka (OZI)-allowed strong decays have been investigated mainly within the $^3P_0$ model~\cite{Lu:2014zua,Yu:2014dda,Godfrey:2015dva,Song:2015nia,Song:2015fha}, the chiral quark model~\cite{Xiao:2014ura,Ni:2021pce}, and Bethe-Salpeter-based approaches~\cite{Li:2017zng}.
However, the conventional quark model has faced a challenge since the discovery of the $D_{s0}^*(2317)^+$ and $D_{s1}(2460)^+$ first reported by the \emph{BABAR}~\cite{BaBar:2003oey} and CLEO~\cite{CLEO:2003ggt} collaborations, respectively.
If the $D_{s0}^*(2317)^+$ and $D_{s1}(2460)^+$ are assigned as the $D_s(1^3P_0)$ and $D_s(1P_1)$ states, their measured masses are far smaller than the conventional quark model expectations~\cite{Godfrey:1985xj,DiPierro:2001dwf}.
Some studies show that the coupled-channel effects are important.
Considering the strong couplings of the bare $D_s(1^3P_0)$ and $D_s(1P_1)$ states to the $S$-wave $DK$ and $D^*K$ channels, respectively, their low mass puzzle can be reasonably explained~\cite{Simonov:2004ar,Rupp:2006sb,Zhou:2011sp,Coito:2011qn,Ortega:2016mms,Yang:2021tvc,Hao:2022vwt,Yang:2023tvc,Ni:2023lvx}.
Thus, to provide more reliable predictions for the higher excitations, it is necessary to carry out a systematic study within an unquenched quark model.

In a recent work~\cite{Ni:2023lvx}, we developed an unquenched quark model associated with chiral dynamics to systematically study the low-lying $1S$, $1P$, $1D$, and $2S$ heavy-light meson states.
In this model, the bare meson states are coupled to the OZI-allowed two meson channels, and the mass shifts arising from the coupled-channel effects, together with the strong decay widths, are calculated with the same strong transition amplitudes described within the chiral quark model.
The masses together with decay widths for all of the well-established resonances can be reasonably explained within the unquenched quark model framework.
The newly observed high-lying state $D_{s1}(2933)^+$ provides a good opportunity to further test our model.
Stimulated by this new observation, in this work, as a continuation of our previous study on the low-lying heavy-light meson spectra~\cite{Ni:2023lvx}, we extend the unquenched quark model calculations of the
mass spectra and OZI-allowed strong decays to the higher $3S$-, $2P$-, $2D$-, and $1F$-wave excitations in the $D$- and $D_s$-meson families.

This paper is organized as follows.
In Sec.~\ref{sec:formalism}, we briefly introduce the unquenched quark model framework.
In Sec.~\ref{sec:results}, we present the calculated mass spectra and strong decay widths, and further discuss the possible assignments for these higher excitations.
Finally, a summary is given in Sec.~\ref{sec:summary}.

\section{Theoretical Framework}\label{sec:formalism}

Following the theoretical framework of our previous studies~\cite{Ni:2021pce, Ni:2023lvx}, the mass spectra and OZI-allowed strong decays of the higher $D$ and $D_s$ mesons are studied within a unified unquenched quark model associated with chiral dynamics.
In the following, we give a brief introduction to this model.

For a bare $Q\bar q$ meson in its rest frame, the Hamiltonian is taken as
\begin{equation}
\label{eq:bare-hamiltonian}
\mathcal{H}_{0}
=
\sqrt{\boldsymbol{p}_1^2+m_1^2}
+
\sqrt{\boldsymbol{p}_2^2+m_2^2}
+
V(r),
\end{equation}
where $m_1$ and $m_2$ are the constituent masses of the light and heavy quarks with momenta $\boldsymbol p_1$ and $\boldsymbol p_2$, respectively, while $r=|\boldsymbol r_1-\boldsymbol r_2|$ is the distance between the two quarks.
The effective interaction is separated into a central potential and a spin-dependent part.
The central potential has the Cornell form
\begin{equation}
\label{eq:central-potential}
V_0(r)
=
-
\frac{4}{3}\frac{\alpha_s(r)}{r}
+
br
+
C_0 .
\end{equation}
The spin-dependent interaction contains the smeared contact term, the tensor force, and the spin-orbit force:
\begin{align}
\label{eq:spin-dependent-potential}
V_{\mathrm{sd}}(r)
&=
\frac{32\pi\alpha_s(r)\sigma^3 e^{-\sigma^2r^2}}
{9\sqrt{\pi^3}\,\tilde m_1m_2}
\boldsymbol S_1\cdot\boldsymbol S_2
+
\frac{4}{3}
\frac{\alpha_s(r)}{\tilde m_1m_2}
\frac{1}{r^3}
S_{12}
+
H_{LS},
\end{align}
with
$
S_{12}
=
\frac{3(\boldsymbol S_1\cdot\boldsymbol r)
(\boldsymbol S_2\cdot\boldsymbol r)}
{r^2}
-
\boldsymbol S_1\cdot\boldsymbol S_2 .
$
The spin-orbit interaction is further separated into symmetric and antisymmetric parts,
\begin{align}
\label{eq:spin-orbit-potential}
H_{LS}
&=
H_{\mathrm{sym}}
+
H_{\mathrm{anti}},
\\
H_{\mathrm{sym}}
&=
\frac{\boldsymbol S_{+}\cdot\boldsymbol L}{2}
\left[
\left(
\frac{1}{2\tilde m_1^2}
+
\frac{1}{2m_2^2}
\right)
G(r)
+
\frac{8\alpha_s(r)}
{3\tilde m_1m_2r^3}
\right],
\\
H_{\mathrm{anti}}
&=
\frac{\boldsymbol S_{-}\cdot\boldsymbol L}{2}
\left(
\frac{1}{2\tilde m_1^2}
-
\frac{1}{2m_2^2}
\right)
G(r),
\end{align}
where $G(r) \equiv 4\alpha_s(r)/(3r^3) - b/r$.
Here $\boldsymbol S_1$ and $\boldsymbol S_2$ denote the spins of the light and heavy quarks, respectively, $\boldsymbol S_{\pm}\equiv\boldsymbol S_1\pm\boldsymbol S_2$, and $\boldsymbol L$ is the relative orbital angular momentum between two quarks.

As in our previous studies~\cite{Ni:2021pce,Ni:2023lvx}, we replace the light quark mass in the spin-dependent terms with an effective mass $\tilde m_1$ to keep the relativistic corrections consistent for the heavy-light systems.
The running coupling is parametrized by $\alpha_s(r)=
\sum_{i=1}^{3}
\alpha_i
\frac{2}{\sqrt{\pi}}
\int_0^{\gamma_i r}
e^{-x^2}\,\mathrm{d}x$~\cite{Godfrey:1985xj}.
In the numerical calculation, we regularize the singular $1/r^3$ terms with the same short range cutoff as in our previous works~\cite{Ni:2021pce,Ni:2023lvx}, and solve Eq.~(\ref{eq:bare-hamiltonian}) with the Gaussian expansion method~\cite{Hiyama:2003cu}.
The radial wave functions are expanded in Gaussian bases, and the masses and wave functions are obtained through the variational method.
With these numerical wave functions, we further calculate the mass shifts and the OZI-allowed strong decay widths.

For heavy-light mesons, the antisymmetric spin-orbit term leads to mixing between the spin-singlet and spin-triplet basis states with $J=L$.
We use the mixing convention as follows
\begin{equation}
\label{eq:state-mixing}
\begin{pmatrix}
nL_J\\
nL'_J
\end{pmatrix}
=
\begin{pmatrix}
\cos\theta_{nL} & \sin\theta_{nL}\\
-\sin\theta_{nL} & \cos\theta_{nL}
\end{pmatrix}
\begin{pmatrix}
n\,{}^1L_J\\
n\,{}^3L_J
\end{pmatrix},
\end{equation}
where $J=L=1,2,3,\cdots$.
With this convention, $nL_J$ and $nL'_J$ denote the lower- and higher-mass eigenstates, respectively.

By solving the Schr\"{o}dinger equation with the Hamiltonian in Eq.~(\ref{eq:bare-hamiltonian}), we first obtain the bare mass $M_A$ and wave function for a $Q\bar q$ state.
By coupling this bare core $|A\rangle$ to the $BC$ continuum components, the physical state $|\Psi(M)\rangle$ is given by~\cite{Kalashnikova:2005ui,Barnes:2007xu,Lu:2016mbb,Ni:2023lvx}
\begin{equation}
\label{eq:physical-state-expansion}
|\Psi(M)\rangle
=
c_A|A\rangle
+
\sum_{BC}
\int
c_{BC}(\boldsymbol q)
|BC,\boldsymbol q\rangle
\,\mathrm{d}^3\boldsymbol q .
\end{equation}
Here, $c_A$ and $c_{BC}(\boldsymbol q)$ are the amplitudes of the bare and continuum components, respectively, and $|BC,\boldsymbol q\rangle$ denotes a two meson continuum state with relative momentum $\boldsymbol q$.
In the coupled space $|A\rangle\oplus |BC,\boldsymbol q\rangle$, the total Hamiltonian is given by
\begin{equation}
	\mathcal{H} = \mathcal{H}_{0} + \mathcal{H}_{c} + \mathcal{H}_{I},
\end{equation}
with $\mathcal H_0|A\rangle=M_A|A\rangle$ and $\mathcal H_c|BC,\boldsymbol q\rangle=E_{BC}(q)|BC,\boldsymbol q\rangle$.
Here, $E_{BC}(q)=\sqrt{M_B^2+q^2}+\sqrt{M_C^2+q^2}$, and $\mathcal H_I$ stands for the interaction which couples the bare state to the continuum channel.
Substituting Eq.~(\ref{eq:physical-state-expansion}) into the Schr\"{o}dinger equation $\mathcal H|\Psi(M)\rangle=M|\Psi(M)\rangle$ and eliminating $c_{BC}(\boldsymbol q)$, we obtain the mass equation~\cite{Kalashnikova:2005ui,Barnes:2007xu,Lu:2016mbb,Ni:2023lvx}
\begin{equation}
	M=M_A+\Sigma(M),
\end{equation}
with
\begin{equation}
\label{eq:self-energy-complex}
\Sigma(M)
=
\sum_{BC}
\int_0^\infty
\frac{
\overline{\left|\mathcal{M}_{A\to BC}(\boldsymbol{q})\right|^2}
}{
M-E_{BC}(q)+i\epsilon
}
q^2\,\mathrm{d}q .
\end{equation}
The averaged transition matrix element in Eq.~(\ref{eq:self-energy-complex}) is defined by 
$\overline{\left|\mathcal{M}_{A\to BC}(\boldsymbol{q})\right|^2}
\equiv
\int
\overline{
	\left|
	\langle BC,\boldsymbol{q}|\mathcal{H}_{I}|A\rangle
	\right|^2
}
\,\mathrm{d}\Omega_{\boldsymbol{q}}$, where the overline denotes the spin average over the initial state and the spin sum over the final states.

The infinitesimal term $i\epsilon$ ($\epsilon\to 0^+$) fixes the analytic continuation above the open channel threshold $M > M_B + M_C$.
The real part of $\Sigma(M)$ gives the mass shift due to the coupled-channel effects, i.e.,
\begin{align}
\label{eq:self-energy}
\Delta M(M)
&=
\operatorname{Re}\Sigma(M) \\
&=
\mathcal{P}
\sum_{BC}
\int_0^\infty
\frac{
\overline{\left|\mathcal{M}_{A\to BC}(\boldsymbol{q})\right|^2}
}{
M-E_{BC}(q)
}
q^2\,\mathrm{d}q .
\end{align}
The imaginary part corresponds to the decay width.
For a two body decay $A\to BC$, one obtains its partial width
\begin{align}
\label{eq:decay-width}
\Gamma(A\to BC)
&=
-2\operatorname{Im}\Sigma_{BC}(M) \\
&=
2\pi
\frac{|\boldsymbol{q}|E_BE_C}{M}
\overline{
\left|
\mathcal{M}_{A\to BC}(\boldsymbol{q})
\right|^2
},
\end{align}
where $\boldsymbol q$ is the on shell relative momentum in the rest frame of $A$.
Thus, the mass shift and the open flavor decay width are determined by the same strong transition matrix element $\mathcal{M}_{A\to BC}(\boldsymbol{q})$.

Following Refs.~\cite{Pennington:2007xr,Zhou:2011sp,Duan:2021alw,Ni:2023lvx,Deng:2023mza,Ni:2025gvx}, the mass shift is calculated with the once subtracted self-energy,
\begin{align}
\label{eq:subtracted-self-energy}
&\Delta M(M,M_0) \nonumber \\
&=
\mathcal{P}
\sum_{BC}
\int_0^\infty
\frac{
(M_0-M)\overline{\left|\mathcal{M}_{A\to BC}(\boldsymbol{q})\right|^2}
}{
[M-E_{BC}(q)][M_0-E_{BC}(q)]
}
q^2\,\mathrm{d}q .
\end{align}
The physical mass $M_{\rm phy}$ is then obtained from $M_{\rm phy}=M_A+\Delta M(M_{\rm phy},M_0)$.
For the present calculation, the subtraction point $M_0$ is chosen as the mass of the corresponding ground state in the $D$ or $D_s$ family, consistent with the approach adopted in our previous work~\cite{Ni:2023lvx}.
This subtraction absorbs the smooth contributions from distant virtual channels into the bare mass, while retaining the energy dependence associated with nearby thresholds. 
For an open channel, the energy denominator changes sign across the on shell momentum, and the contributions on the two sides of the pole partially cancel. 
Therefore, the mass shift does not simply increase with the number of open channels, but depends on the channel thresholds and transition amplitudes of each state.

The transition matrix elements are calculated with $\mathcal H_I$ in the chiral quark model.
The chiral quark model has been successfully applied to describe strong decays of heavy-light mesons~\cite{Zhong:2008kd,Zhong:2009sk,Zhong:2010vq,Xiao:2014ura,li:2021hss,Ni:2021pce} and baryons~\cite{Zhong:2007gp,Liu:2012sj,Wang:2017kfr,Xiao:2017udy,Wang:2017hej,Wang:2018fjm,Wang:2020gkn}.
For light pseudoscalar meson emission, the quark-meson interaction is taken as~\cite{Manohar:1983md,Li:1994cy,Li:1997gd,Zhao:2002id}
\begin{equation}
\label{eq:chiral-lagrangian}
\mathcal{L}_{P}
=
\sum_j
\frac{1}{f_m}
\bar{\psi}_j
\gamma^j_{\mu}
\gamma^j_{5}
\psi_j
\vec{\tau}\cdot\partial^{\mu}\vec{\phi}_m ,
\end{equation}
where $\psi_j$ is the $j$th light quark field, $\phi_m$ denotes the emitted pseudoscalar field, $\tau$ is the isospin operator, and $f_m$ is the corresponding decay constant.
After a Foldy-Wouthuysen reduction~\cite{Arifi:2021orx,Arifi:2022ntc} of Eq.~(\ref{eq:chiral-lagrangian}), the transition operator is written as
\begin{equation}
	\mathcal H_I=\mathcal H_I^{\mathrm{NR}}+\mathcal H_I^{\mathrm{RC}},
\end{equation}
where $\mathcal H_I^{\mathrm{NR}}$ is the leading nonrelativistic operator and $\mathcal H_I^{\mathrm{RC}}$ collects the relativistic correction terms kept up
to order $1/m^2$~\cite{Ni:2023lvx}. 
Explicitly,
\begin{equation}
	\label{eq:transition-operator-nr}
	\mathcal{H}_{I}^{\mathrm{NR}}
	=
	g
	\sum_j
	\left[
	\mathcal{G}\,
	\boldsymbol{\sigma}_{j}\cdot\boldsymbol{q}
	+
	\frac{\omega_m}{2\mu_q}
	\boldsymbol{\sigma}_{j}\cdot\boldsymbol{p}_{j}
	\right]
	I_j\varphi_m ,
\end{equation}
and
\begin{align}
	\label{eq:transition-operator-rc}
	\mathcal{H}_{I}^{\mathrm{RC}}
	&=
	-
	\frac{g}{32\mu_q^2}
	\sum_j
	\Big[
	m_{\mathbb{P}}^2
	\boldsymbol{\sigma}_{j}\cdot\boldsymbol{q}
	\nonumber \\
	&\qquad 
	+
	2\boldsymbol{\sigma}_{j}\cdot
	\left(
	\boldsymbol{q}-2\boldsymbol{p}_{j}
	\right)
	\times
	\left(
	\boldsymbol{q}\times\boldsymbol{p}_{j}
	\right)
	\Big]
	I_j\varphi_m .
\end{align}
Here $\boldsymbol p_j$ and $\boldsymbol\sigma_j$ are the internal momentum and spin operators of the active light quark; $\boldsymbol q$ and $\omega_m$ are the three momentum and energy of the emitted pseudoscalar meson in the initial meson rest frame.
The plane wave factor is $\varphi_m=\exp(-i\boldsymbol q\cdot\boldsymbol r_j)$, and $I_j$ is the flavor operator in SU(3) flavor space~\cite{Li:1997gd}.
The emitted pseudoscalar meson mass is denoted by $m_{\mathbb P}$.
The factors in Eqs.~(\ref{eq:transition-operator-nr}) and~(\ref{eq:transition-operator-rc}) are $g=\frac{\delta}{f_m}\sqrt{(E_i+M_i)(E_f+M_f)}$ and $\mathcal{G}=-\left(1+\frac{\omega_m}{E_f+M_f}+\frac{\omega_m}{2m'_j}\right)$.
Here $M_i$ and $M_f$ are the initial and final heavy-light meson masses, $E_i$ and $E_f$ are their energies in the chosen decay frame, and $m_j$ and $m'_j$ are the constituent masses of the active light quark before and after the pseudoscalar emission.
The reduced mass parameter is defined by $1/\mu_q=1/m_j+1/m'_j$.

For channels containing a light vector meson, we adopt the quark vector meson coupling introduced in Ref.~\cite{Ni:2021pce}.
In the SU(3) flavor basis, this interaction is taken as~\cite{Zhao:1998fn,Zhao:2000tb}
\begin{equation}
\label{eq:vector-lagrangian}
\mathcal{L}_{V}
=
\sum_j
\bar{\psi}_j
\left(
a\gamma^{j}_{\mu}
+
\frac{i b}{2m_j}
\sigma_{\mu\nu}q^\nu
\right)
V^\mu
\psi_j ,
\end{equation}
where $V^\mu$ is the emitted vector meson field, and $a$ and $b$ are the vector and tensor coupling strengths, respectively.
After the nonrelativistic reduction, the transverse and longitudinal transition operators are~\cite{Zhao:1998fn}
\begin{align}
\label{eq:vector-transition-transverse}
\mathcal{H}_{I}^{V,T}
&=
\sum_j
\left[
i\frac{b'}{2m_q}
\boldsymbol{\sigma}_j\cdot
\left(
\boldsymbol q\times\boldsymbol\epsilon
\right)
+
\frac{a}{2\mu_q}
\boldsymbol p_j\cdot\boldsymbol\epsilon
\right]
I_j\varphi_m ,
\end{align}
\begin{align}
\mathcal{H}_{I}^{V,L}
&=
\sum_j
\frac{aM_V}{|\boldsymbol q|}
I_j\varphi_m .\label{eq:vector-transition-longitudinal}
\end{align}
Here $\boldsymbol\epsilon$ is the transverse polarization vector, $M_V$ is the vector meson mass, $m_q$ is the light constituent quark mass used in the vector vertex, and $b'\equiv b-a$.

Since the effective chiral quark-meson interaction is valid in the low momentum region, the pointlike vertices may introduce unphysical contributions at high momenta in the loop integrals. 
To suppress these high momentum contributions, following Ref.~\cite{Ni:2023lvx} we introduce a Gaussian form factor into the transition amplitude,
\begin{equation}
\label{eq:gaussian-suppression}
\langle BC,\boldsymbol{q}|\mathcal{H}_{I}|A\rangle
\rightarrow
\langle BC,\boldsymbol{q}|
\mathcal{H}_{I}
e^{-\frac{q^2}{2\Lambda^2}}
|A\rangle .
\end{equation}
With the cutoff parameter $\Lambda$, the mass shifts mainly arise from the low-momentum region ($q \lesssim \Lambda$), while for open decay channels the suppression factor is $e^{-q_i^2/\Lambda^2}$ at the on-shell momentum $q_i$.
Similar suppression factors have been widely used in previous coupled-channel calculations~\cite{Silvestre-Brac:1991qqx,Ortega:2016pgg,Ortega:2016mms,Yang:2021tvc,Yang:2022vdb}.

The parameters in the model Hamiltonian are directly taken from Ref.~\cite{Ni:2023lvx}.
In Table II of Ref.~\cite{Ni:2023lvx}, two parameter sets are given.
One set is for calculations of the mass spectrum in the quenched framework.
The other set is for calculations of the mass spectrum within the unquenched framework. 
With these parameters, the Hamiltonian gives the quenched reference masses $M_{\rm Q}$, while the corresponding bare masses $M_A$ and wave functions are used in the unquenched calculation.
For the strong transition amplitude calculations, the parameters related to the pseudoscalar meson-quark coupling are also fixed as in Ref.~\cite{Ni:2023lvx}: $\delta=0.557$, $f_\pi=132~\mathrm{MeV}$, $f_K=f_\eta=160~\mathrm{MeV}$, $\mu_q=225~\mathrm{MeV}$, and $\Lambda=0.78~\mathrm{GeV}$.
For the decay channels containing a light vector meson, we take $a=-1.7$ and $b'=2.5$, which were extracted from the $\omega$ and $\rho$ photoproduction processes by Zhao \textit{et al.}~\cite{Zhao:1998rt,Zhao:1998fn}.

The effective chiral quark-meson interaction is valid in the low momentum region below the chiral symmetry breaking scale $\Lambda_{\chi{\rm SB}}=4\pi f_m \sim 1~\mathrm{GeV}$~\cite{Manohar:1983md}. 
To suppress the unphysical high momentum contributions, the Gaussian cutoff $\Lambda$ is introduced around the chiral scale with a typical range of $0.8\pm0.2~\mathrm{GeV}$.
In our previous study of the low-lying heavy-light meson spectra~\cite{Ni:2023lvx}, the cutoff $\Lambda=0.78~\mathrm{GeV}$ was determined by reproducing the physical masses of $D_{s0}^*(2317)$ and $D_{s1}(2460)$ after including coupled-channel effects. 
With this cutoff, our model also reasonably described the masses and strong decay widths of other established charmed and charmed-strange mesons. 
We retain this value for the higher excitations without readjusting it to the present spectrum.

The meson masses used in the calculations are taken from the Review of Particle Physics (RPP)~\cite{ParticleDataGroup:2024cfk} for the well-established states; for the remaining states, our predicted masses are adopted.
It should be emphasized that the main potential model parameters were fixed by the low-lying spectrum in our previous work~\cite{Ni:2023lvx}, and the remaining inputs were adopted as specified above, without readjustment to the higher states. Details on the determination of the potential model parameters and other parameters can be found in its Appendix. The results presented below are therefore genuine theoretical predictions within this unified framework.
With these parameter sets, the quenched and unquenched mass spectra and the OZI-allowed strong decay widths of the $3S$-, $2P$-, $2D$-, and $1F$-wave $D$ and $D_s$ mesons can be obtained.

\begin{table*}[t]
\caption{Mass spectra of the higher excited charmed and charmed-strange mesons in the quenched (Q) and unquenched (UQ) quark models. Here, $M_{\rm Q}$ is the quenched mass, while $M_A$, $\Delta M$, and $M_{\rm phy}$ denote the unquenched bare mass, the diagonal mass shift induced by coupled-channel effects, and the corresponding physical mass, respectively. The unit of mass is MeV.}
\label{tab:higher_ccef_spectrum}
\small
\setlength{\tabcolsep}{2.0pt}
\renewcommand{\arraystretch}{1.03}
\begin{tabular*}{\textwidth}{@{\extracolsep{\fill}}lcccccccccccc@{}}
\hline \hline
\multirow{2}{*}{State} & \multirow{2}{*}{$J^P$} & \multirow{2}{*}{$M_{\rm Q}$} & \multicolumn{3}{c}{Unquenched} & \multirow{2}{*}{\shortstack{Experiment}} & \multirow{2}{*}{GM~\cite{Godfrey:2015dva}} & \multirow{2}{*}{EFG~\cite{Ebert:2009ua}} & \multirow{2}{*}{NLZ~\cite{Ni:2021pce}} & \multirow{2}{*}{ZVR~\cite{Zeng:1994vj}} & \multirow{2}{*}{LJM~\cite{Li:2010vx}} & \multirow{2}{*}{LNR~\cite{Lahde:1999ih}} \\
\cline{4-6}
 &  &  & $M_A$ & $\Delta M$ & $M_{\rm phy}$ &  &  &  &  &  &  &  \\
\hline
\multicolumn{13}{l}{\textit{$D$ mesons}} \\
$D(3^1S_0)$ & $0^-$ & 3036 & 3071 & -147 & 2924 & $\cdots$ & 3068 & 3062 & 3029 & 2980 & $\cdots$ & 2904 \\
$D(3^3S_1)$ & $1^-$ & 3098 & 3136 & -56 & 3080 & $\cdots$ & 3110 & 3096 & 3093 & 3070 & $\cdots$ & 2947 \\
$D(2^3P_0)$ & $0^+$ & 2849 & 2871 & -198 & 2673 & $\cdots$ & 2931 & 2919 & 2849 & 2780 & 2752 & 2758 \\
$D(2P_1)$ & $1^+$ & 2902 & 2937 & -162 & 2775 & $\cdots$ & 2924 & 2932 & 2900 & 2890 & 2886 & 2792 \\
$D(2P_1^\prime)$ & $1^+$ & 2938 & 2970 & -140 & 2830 & $\cdots$ & 2961 & 3021 & 2936 & 2890 & 2926 & 2802 \\
$D(2^3P_2)$ & $2^+$ & 2958 & 3002 & -102 & 2900 & $\cdots$ & 2957 & 3012 & 2955 & 2940 & 2971 & 2860 \\
$D(2^3D_1)$ & $1^-$ & 3144 & 3168 & -205 & 2963 & $\cdots$ & 3231 & 3228 & 3143 & 3130 & 3168 & 3052 \\
$D(2D_2)$ & $2^-$ & 3171 & 3217 & -148 & 3069 & $\cdots$ & 3212 & 3259 & 3168 & 3160 & 3145 & 2997 \\
$D(2D'_2)$ & $2^-$ & 3221 & 3260 & -104 & 3156 & $\cdots$ & 3248 & 3307 & 3221 & 3170 & 3215 & 3029 \\
$D(2^3D_3)$ & $3^-$ & 3204 & 3257 & -81 & 3176 & $\cdots$ & 3226 & 3335 & 3202 & 3190 & 3170 & 2999 \\
$D(1^3F_2)$ & $2^+$ & 3093 & 3125 & -135 & 2990 & $\cdots$ & 3132 & 3090 & 3096 & 3000 & $\cdots$ & $\cdots$ \\
$D(1F_3)$ & $3^+$ & 3027 & 3078 & -113 & 2965 & $\cdots$ & 3108 & 3129 & 3022 & 3010 & $\cdots$ & $\cdots$ \\
$D(1F_3')$ & $3^+$ & 3122 & 3167 & -88 & 3079 & $\cdots$ & 3143 & 3145 & 3129 & 3030 & $\cdots$ & $\cdots$ \\
$D(1^3F_4)$ & $4^+$ & 3035 & 3092 & -78 & 3014 & $\cdots$ & 3113 & 3187 & 3034 & 3030 & $\cdots$ & $\cdots$ \\
\hline\hline
\multicolumn{13}{l}{\textit{$D_s$ mesons}} \\
\multirow{2}{*}{$D_s(3^1S_0)$} & \multirow{2}{*}{$0^-$} & \multirow{2}{*}{3132} & \multirow{2}{*}{3160} & \multirow{2}{*}{-71} & \multirow{2}{*}{3089} & $D_{sJ}(3040)^+?$ & \multirow{2}{*}{3154} & \multirow{2}{*}{3219} & \multirow{2}{*}{3126} & \multirow{2}{*}{3090} & \multirow{2}{*}{$\cdots$} & \multirow{2}{*}{3044} \\
 &  &  &  &  &  & $3044\pm 8^{+30}_{-5}$ &  &  &  &  &  &  \\
$D_s(3^3S_1)$ & $1^-$ & 3193 & 3213 & -40 & 3173 & $\cdots$ & 3193 & 3242 & 3191 & 3190 & $\cdots$ & 3087 \\
$D_s(2^3P_0)$ & $0^+$ & 2939 & 2944 & -80 & 2864 & $\cdots$ & 3005 & 3054 & 2940 & 2900 & 2830 & 2901 \\
\multirow{2}{*}{$D_s(2P_1)$} & \multirow{2}{*}{$1^+$} & \multirow{2}{*}{3000} & \multirow{2}{*}{3019} & \multirow{2}{*}{-47} & \multirow{2}{*}{2972} & $D_{s1}(2933)^+?$ & \multirow{2}{*}{3018} & \multirow{2}{*}{3067} & \multirow{2}{*}{3002} & \multirow{2}{*}{3000} & \multirow{2}{*}{2958} & \multirow{2}{*}{2928} \\
 &  &  &  &  &  & $2933^{+6+4}_{-5-3}$ &  &  &  &  &  &  \\
$D_s(2P_1^\prime)$ & $1^+$ & 3028 & 3044 & -40 & 3004 & $\cdots$ & 3038 & 3154 & 3026 & 3010 & 2995 & 2942 \\
$D_s(2^3P_2)$ & $2^+$ & 3054 & 3083 & -44 & 3039 & $\cdots$ & 3048 & 3142 & 3053 & 3060 & 3040 & 2988 \\
$D_s(2^3D_1)$ & $1^-$ & 3232 & 3234 & -69 & 3165 & $\cdots$ & 3306 & 3383 & 3233 & 3250 & 3217 & 3172 \\
$D_s(2D_2)$ & $2^-$ & 3267 & 3293 & -53 & 3240 & $\cdots$ & 3298 & 3403 & 3267 & 3280 & 3217 & 3144 \\
$D_s(2D'_2)$ & $2^-$ & 3306 & 3325 & -54 & 3271 & $\cdots$ & 3323 & 3456 & 3306 & 3290 & 3260 & 3167 \\
$D_s(2^3D_3)$ & $3^-$ & 3297 & 3332 & -44 & 3288 & $\cdots$ & 3311 & 3469 & 3299 & 3310 & 3240 & 3157 \\
$D_s(1^3F_2)$ & $2^+$ & 3174 & 3182 & -28 & 3154 & $\cdots$ & 3208 & 3230 & 3176 & 3120 & $\cdots$ & $\cdots$ \\
$D_s(1F_3)$ & $3^+$ & 3124 & 3152 & -24 & 3128 & $\cdots$ & 3186 & 3254 & 3123 & 3130 & $\cdots$ & $\cdots$ \\
$D_s(1F_3')$ & $3^+$ & 3200 & 3224 & -48 & 3176 & $\cdots$ & 3218 & 3266 & 3205 & 3150 & $\cdots$ & $\cdots$ \\
$D_s(1^3F_4)$ & $4^+$ & 3131 & 3166 & -41 & 3125 & $\cdots$ & 3190 & 3300 & 3134 & 3160 & $\cdots$ & $\cdots$ \\
\hline \hline
\end{tabular*}
\end{table*}

\begin{figure*}[t]
	\centering
	\includegraphics[width=1.00\textwidth]{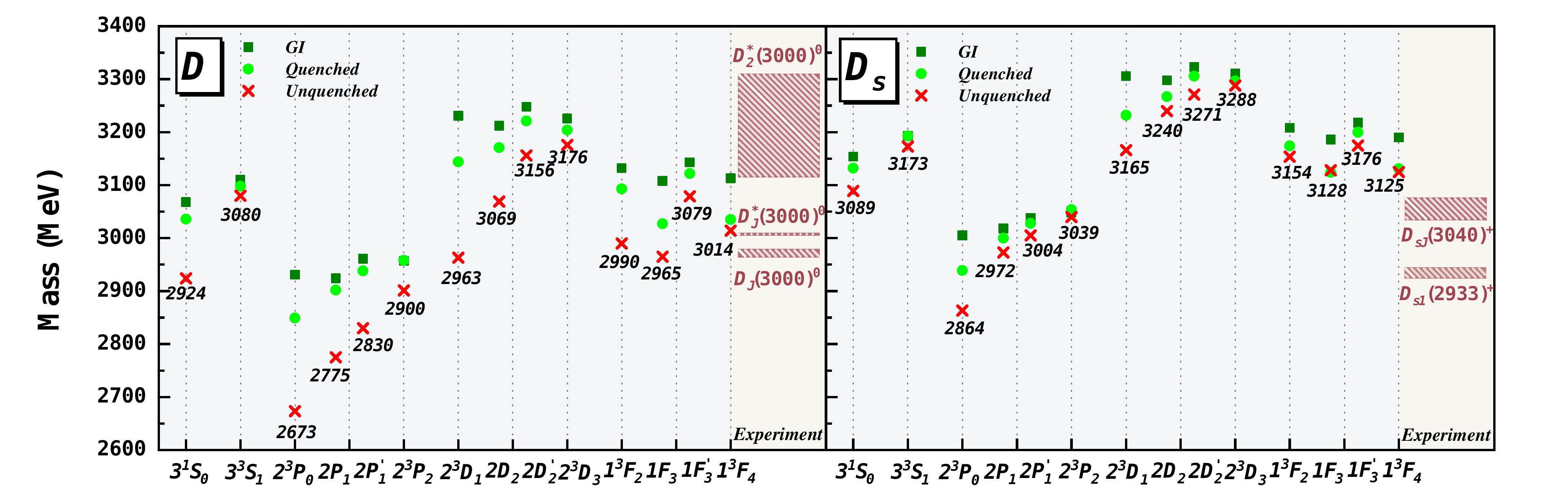}
	\caption{Mass spectra of $3S$-, $2P$-, $2D$-, and $1F$-wave $D$ and $D_s$ mesons. The green circles and red crosses denote the quenched masses $M_{\rm Q}$ and the dressed masses $M_{\rm phy.}$ after including coupled-channel effects, while the dark green squares show the Godfrey-Isgur (GI) model predictions~\cite{Godfrey:2015dva}. The numbers beside the red crosses are the calculated dressed masses in MeV. The red-hatched bands mark the experimental mass regions of the observed structures used for comparison.}
	\label{fig:mass_spectra}
\end{figure*}

\section{Results and discussion}
\label{sec:results}

The obtained mass spectra of the higher $3S$-, $2P$-, $2D$-, and $1F$-wave excitations in the $D$- and $D_s$-meson families are listed in Table~\ref{tab:higher_ccef_spectrum} and shown in Fig.~\ref{fig:mass_spectra}.
Moreover, the channel contributions to the mass shifts and OZI-allowed strong decay widths are listed in Tables~\ref{tab:D_3S_channel_ccef}--\ref{tab:Ds_1F_channel_ccef} as an Appendix.

For higher excited states, an increasing number of decay channels open up as the excitation energy increases. 
One might naturally expect that states with more open channels would acquire larger coupled-channel mass shifts and broader total widths. 
However, an inspection of these tables reveals that neither the mass shift nor the total width simply scales with the number of open channels. 
Both quantities depend strongly on the decay partial waves, with the decay width further governed by the phase space, and the mass shift sensitive to the distance to each channel threshold. 
Furthermore, for the radial excitations, the radial nodes in the wave functions can lead to cancellations in the transition matrix elements, thereby affecting both the mass shifts and decay widths. 
A clear example is the comparison between $D_s(2^3P_0)$ and $D_s(2^3D_3)$. 
For $D_s(2^3P_0)$, although only two channels are open, its decays proceed through $S$ waves and its mass is close to the thresholds, which yields a large width $\Gamma_{\rm tot}\simeq 201~\mathrm{MeV}$ and mass shift $\Delta M\simeq -80~\mathrm{MeV}$. 
In contrast, for $D_s(2^3D_3)$, although 23 channels are open, its dominant decays proceed through $F$ waves with a strong centrifugal barrier, leading to a much smaller width of $\Gamma_{\rm tot}\simeq 31~\mathrm{MeV}$ and mass shift of $\Delta M\simeq -44~\mathrm{MeV}$.

In the $D$-meson sector, after including the mass shifts induced by coupled-channel effects, we obtain the following physical mass ranges for the higher $3S$, $2P$, $2D$, and $1F$ excitations:
\begin{align}
	M_{\rm phy}[D(3S)]&\simeq2.92\text{--}3.08~\mathrm{GeV},\nonumber\\
	M_{\rm phy}[D(2P)]&\simeq2.67\text{--}2.90~\mathrm{GeV},\nonumber\\
	M_{\rm phy}[D(2D)]&\simeq2.96\text{--}3.18~\mathrm{GeV},\nonumber\\
	M_{\rm phy}[D(1F)]&\simeq2.97\text{--}3.08~\mathrm{GeV}.
\end{align}
Several quenched quark model results are collected in Table~\ref{tab:higher_ccef_spectrum} for comparison.
More predictions for the $D$-meson spectrum can be found in Refs.~\cite{Godfrey:1985xj,Zhang:1993ik,Ebert:1997nk,DiPierro:2001dwf,Vijande:2006hj,Badalian:2011tb,Lu:2014zua,Song:2015fha,Gandhi:2019lta}.
It is found that the coupled-channel effects are sizeable for several higher $D$-meson states.
Compared with the quenched quark model expectations~\cite{Ebert:2009ua,Godfrey:2015dva,Ni:2021pce}, the physical masses of $D(3^1S_0)$, $D(2^3P_0)$, $D(2P_1)$, $D(2P_1^\prime)$, $D(2^3D_1)$, $D(2D_2)$, and $D(1^3F_2)$ are shifted to lower masses by about $100$--$270~\mathrm{MeV}$ in the present calculation.
A recent coupled-channel study of the $D$-meson spectrum within the $^3P_0$ model~\cite{Hao:2024ptu} also obtains significant mass shifts.
However, the physical masses obtained in Ref.~\cite{Hao:2024ptu} are generally higher than ours, which may be related to the different bare spectra, coupled-channel scheme, and transition operators adopted in the two calculations.

In the $D_s$-meson sector, after including the mass shifts induced by coupled-channel effects, we obtain the following physical mass ranges:
\begin{align}
	M_{\rm phy}[D_s(3S)]&\simeq3.09\text{--}3.17~\mathrm{GeV},\nonumber\\
	M_{\rm phy}[D_s(2P)]&\simeq2.86\text{--}3.04~\mathrm{GeV},\nonumber\\
	M_{\rm phy}[D_s(2D)]&\simeq3.17\text{--}3.29~\mathrm{GeV},\nonumber\\
	M_{\rm phy}[D_s(1F)]&\simeq3.13\text{--}3.18~\mathrm{GeV}.
\end{align}
Several quenched quark model results are collected in Table~\ref{tab:higher_ccef_spectrum} for comparison.
For the $D_s$ mesons, the mass shifts induced by coupled-channel effects range from $-23.8$ to $-80.2~\mathrm{MeV}$. 
Sizable downward shifts are obtained for $D_s(2^3P_0)$, $D_s(3^1S_0)$, and $D_s(2^3D_1)$, with values of $-80.2$, $-71.3$, and $-68.8~\mathrm{MeV}$, respectively, while those for the other states are relatively moderate, ranging from $-24$ to $-54~\mathrm{MeV}$. 
With these mass shifts, the calculated physical mass of the low mass state $D_s(2P_1)$ ($2972~\mathrm{MeV}$) is close to the newly observed $D_{s1}(2933)^+$, while that of $D_s(3^1S_0)$ ($3089~\mathrm{MeV}$) lies near the broad structure $D_{sJ}(3040)^+$. 
To test these possible assignments, their decay widths and dominant decay modes will be analyzed in detail in the following subsections.

\subsection{$D_{s1}(2933)^+$}

We first examine whether the newly observed $D_{s1}(2933)^+$ can be assigned as one of the axial vector states in the $D_s$-meson family.
The LHCb collaboration observed this state through a full amplitude analysis of $B^0\to D^+D^-K^+\pi^-$ decays with a statistical significance exceeding $10\sigma$~\cite{LHCb:2026sup}.
Its Breit-Wigner mass and width are $M_{\rm exp}=2933^{+6+4}_{-5-3}~\mathrm{MeV}$ and $\Gamma_{\rm exp}=72^{+18+7}_{-12-10}~\mathrm{MeV}$, and the spin-parity is determined to be $J^P=1^+$~\cite{LHCb:2026sup}.
In the amplitude fit, the $D_{s1}(2933)^+$ is reconstructed in the $D^+K^+\pi^-$ system, and its dominant contributions come from the intermediate states $D^+K^*(892)^0$ and $K^+D_2^*(2460)^0$~\cite{LHCb:2026sup}.

With the determined quantum number $J^P=1^+$, the $D_{s1}(2933)^+$ can be assigned as either the low mass state $D_s(2P_1)$ or the high mass state $D_s(2P_1^\prime)$.
The masses of the two $D_s(2P)$ axial vector states are predicted to be in the range of about $2.96$--$3.26~\mathrm{GeV}$ in the literature~\cite{Li:2010vx,Godfrey:2015dva,Ni:2021pce,Hao:2022vwt,Yang:2023tvc}, with a model-dependent splitting of about $20$--$230~\mathrm{MeV}$.
The measured mass of the $D_{s1}(2933)^+$ is closer to the low mass state $D_s(2P_1)$.
However, the mass value alone cannot distinguish these two possibilities.
The decay width together with the dominant decay modes must be further examined.

The strong decays of the two $D_s(2P)$ axial vector states have been extensively studied in the $^3P_0$ model~\cite{Sun:2009tg,Li:2010vx,Song:2015nia,Godfrey:2015dva,Segovia:2015dia,Jiang:2024ftx}, the chiral quark model~\cite{Zhong:2009sk,Ni:2021pce}, and coupled-channel calculations~\cite{Hao:2022vwt,Yang:2023tvc}.
In these calculations, the low mass state $D_s(2P_1)$ is predicted to be a broad state with a total width of $\Gamma\simeq140$--$400~\mathrm{MeV}$~\cite{Zhong:2009sk,Li:2010vx,Song:2015nia,Godfrey:2015dva,Ni:2021pce,Hao:2022vwt,Jiang:2024ftx}.
In several quark model calculations, the high mass state $D_s(2P_1^\prime)$ is relatively narrow, with $\Gamma\simeq120$--$160~\mathrm{MeV}$~\cite{Li:2010vx,Song:2015nia,Godfrey:2015dva,Ni:2021pce}.
It should be mentioned that the measured width of the $D_{s1}(2933)^+$, $\Gamma_{\rm exp}=72^{+18+7}_{-12-10}~\mathrm{MeV}$, is significantly narrower than the previous predictions for the low mass state.

To further test the $D_s(2P_1)$ assignment for the $D_{s1}(2933)^+$, we first calculate its mass in the unquenched framework.
For the low mass state $D_s(2P_1)$, the bare mass $M_A=3019~\mathrm{MeV}$ lies above the measured mass of the $D_{s1}(2933)^+$.
After including coupled-channel effects, we obtain a mass shift $\Delta M=-47~\mathrm{MeV}$ and a physical mass
\begin{align}
	M_{\rm phy}\simeq2972~\mathrm{MeV},
\end{align}
which is close to the observed $D_{s1}(2933)^+$.
It is found that this mass shift mainly arises from the $D_2^*(2460)K$ channel ($\Delta M_i\simeq-25~\mathrm{MeV}$) because its threshold lies close to the bare mass of $D_s(2P_1)$.
%Furthermore, the two $D_s(2P)$ axial vector states share several open channels, which can induce additional mixing through meson loops beyond the $^3P_1$-$^1P_1$ mixing caused by the antisymmetric spin-orbit interaction.
%Using the framework of Ref.~\cite{Ni:2025gvx}, we find that this off-diagonal mixing shifts the masses by only a few MeV and does not affect the present assignment.

In addition to the mass, the decay width provides another important observable to test this assignment.
As shown in Table~\ref{tab:Ds_2P_channel_ccef}, our predicted total width for the $D_s(2P_1)$ state is
\begin{align}
	\Gamma\simeq92~\mathrm{MeV},
\end{align}
which is much narrower than the previous predictions~\cite{Zhong:2009sk,Li:2010vx,Song:2015nia,Godfrey:2015dva,Ni:2021pce,Hao:2022vwt,Jiang:2024ftx}, and is compatible with the measured width $\Gamma_{\rm exp}=72^{+18+7}_{-12-10}~\mathrm{MeV}$.
This relatively narrow width obtained in the present work is mainly caused by two effects.
First, the coupled-channel effects move the physical state closer to the $D^*K$ threshold, which reduces the phase space for this dominant decay channel. 
Second, an additional Gaussian factor $e^{-q^2/(2\Lambda^2)}$ is introduced in the transition operator to soften the hard vertex of quark pair creation in vacuum, which suppresses the nonphysical contributions in the high momentum region.
With these two effects, the calculated width of the $D_s(2P_1)$ state becomes compatible with the measured width of the $D_{s1}(2933)^+$.

Because the decays are sensitive to the phase space, we recalculate the partial widths of the $D_s(2P_1)$ state with the measured mass $M_{\rm exp}=2933~\mathrm{MeV}$.
The results are listed in Table~\ref{tab:Ds_2P_exp_widths}.
With the experimental mass, the obtained total width is
\begin{align}
	\Gamma \simeq80~\mathrm{MeV},
\end{align}
which is in good agreement with the $D_{s1}(2933)^+$ measurement~\cite{LHCb:2026sup}.
Thus, both the calculated mass and width favor assigning the $D_{s1}(2933)^+$ to the low mass state $D_s(2P_1)$.
For this assignment, the state mainly decays into the $D^*K$ channel with a branching fraction of $\sim 67\%$.
Meanwhile, the $DK^*$ channel provides a sizable contribution of $\sim 20\%$, giving a partial width ratio
\begin{align}
	R=\frac{\Gamma(DK^*)}{\Gamma(D^*K)}\simeq0.3.
\end{align}
This $DK^*$ mode includes the charged $D^+K^*(892)^0$ component observed in the LHCb amplitude analysis~\cite{LHCb:2026sup}.
The predicted ratio $R\simeq0.3$ can be used to further test this assignment in future experiments.

\begin{table}[t]
	\caption{Partial decay widths of the two $D_s(2P)$ axial vector states. 
		The low mass state $D_s(2P_1)$ is calculated at the central mass of the $D_{s1}(2933)^+$, while the high mass state $D_s(2P_1^\prime)$ is calculated at $M_{\rm exp}=3044~\mathrm{MeV}$ for comparison with the $D_{sJ}(3040)^+$ assignment. 
		The unit is MeV.}
	\label{tab:Ds_2P_exp_widths}
	\small
	\setlength{\tabcolsep}{2.5pt}
	\renewcommand{\arraystretch}{1.02}
	\begin{tabular*}{\columnwidth}{@{\extracolsep{\fill}}lcc@{}}
		\hline \hline
		& $D_s(2P_1)$             & $D_s(2P_1^\prime)$   \\
		Channel & as $D_{s1}(2933)^+$     & as $D_{sJ}(3040)^+$ \\
		\hline
		$D^*K$ & 53.7 & 30.8 \\
		$D_0^*(2300)K$ & 0.1 & 24.9 \\
		$D_1(2430)K$ & 1.5 & 0.8 \\
		$D_1(2420)K$ & $4.4\times 10^{-2}$ & 35.5 \\
		$D_2^*(2460)K$ & $\cdots$ & 8.8 \\
		$D_s^*\eta$ & 8.7 & 6.4 \\
		$D_{s0}^*(2317)\eta$ & $1.5\times 10^{-2}$ & 2.6 \\
		$D_{s1}(2460)\eta$ & $\cdots$ & $7.5\times 10^{-2}$ \\
		$DK^*$ & 15.8 & 0.2 \\
		$D^*K^*$ & $9.4\times 10^{-2}$ & 3.1 \\
		$D_s\phi$ & $\cdots$ & 0.3 \\
		\hline
		$\mathrm{Total}$ & 79.8 & 113.4 \\
		\hline\hline
	\end{tabular*}
\end{table}

\subsection{$D_{sJ}(3040)^+$}

In the $D_s$-meson family, the broad structure $D_{sJ}(3040)^+$ was observed by the \emph{BABAR} collaboration in the $D^*K$ invariant mass spectrum from inclusive $e^+e^-$ production~\cite{BaBar:2009rro}.
Its measured mass and width are $M_{\rm exp}=3044\pm8^{+30}_{-5}~\mathrm{MeV}$ and $\Gamma_{\rm exp}=239\pm35^{+46}_{-42}~\mathrm{MeV}$, respectively.
Although its spin-parity has not been determined, the absence of a $DK$ signal and the helicity angle distribution in the $D^*K$ channel favor an unnatural parity assignment with possible $J^P=0^-, 1^+,2^-$, and so on~\cite{BaBar:2009rro}.
Furthermore, the LHCb collaboration found weak evidence for a structure around $3040~\mathrm{MeV}$ in the $D^*K$ final state, which is also compatible with unnatural parity contributions~\cite{LHCb:2016mbk}.

Theoretically, this broad structure has been widely investigated by combining the spectrum and decay properties within various models~\cite{Chen:2009zt,Sun:2009tg,Jiang:2024ftx,Zhong:2009sk,Badalian:2011tb,Godfrey:2015dva,Song:2015nia,Li:2017zng,
Ni:2021pce,Xu:2014mqa,Hao:2022vwt,Yang:2023tvc}.
Most of these studies support assigning the $D_{sJ}(3040)^+$ as one of the $D_s(2P)$ axial vector states.
However, its specific identification with the low mass state $D_s(2P_1)$ or the high mass state $D_s(2P_1^\prime)$ remains
model dependent.
Meanwhile, effective Lagrangian analyses based on heavy quark symmetry suggest that other unnatural parity assignments cannot be excluded, and the partial width ratio $\Gamma(D_s^*\eta)/\Gamma(D^*K)$ may provide a useful criterion~\cite{Colangelo:2010te}.
To further test these possibilities, we compare the $D_s(2P_1^\prime)$ and $D_s(3^1S_0)$ assignments for the $D_{sJ}(3040)^+$ within our coupled-channel model.

For the $D_s(2P_1^\prime)$, our predicted physical mass is
\begin{align}
	M_{\rm phy}\simeq3004~\mathrm{MeV},
\end{align}
which lies about $30~\mathrm{MeV}$ above the low-mass state $D_s(2P_1)$.
Although this mass is comparable to the $D_{sJ}(3040)^+$ measurement, the predicted total width $\Gamma\simeq96~\mathrm{MeV}$ is much smaller than the experimental value $\Gamma_{\rm exp}=239\pm35^{+46}_{-42}~\mathrm{MeV}$~\cite{BaBar:2009rro}.
Even with the measured mass $M_{\rm exp}=3044~\mathrm{MeV}$ as an input, our predicted total width,
\begin{align}
	\Gamma \simeq113~\mathrm{MeV},
\end{align}
is still too narrow to be comparable with the data.
Furthermore, as listed in Table~\ref{tab:Ds_2P_exp_widths}, the main decay channels are $D_1(2420)K$, $D^*K$, and $D_0^*(2300)K$, with branching fractions of $\sim 31\%$, $\sim 27\%$, and $\sim 22\%$, respectively.
If the $D_{sJ}(3040)^+$ is indeed the $D_s(2P_1^\prime)$, it should be established in the $D_1(2420)K$ final state as well.
To establish the high mass axial vector state $D_s(2P_1^\prime)$ and clarify the nature of $D_{sJ}(3040)^+$, the $D_1(2420)K$ channel is worth observing in future experiments.

Another possible assignment of $D_{sJ}(3040)^+$ is the pseudoscalar resonance $D_s(3^1S_0)$.
In the conventional quark model based on the quenched approximation, the predicted mass is often in the range of $3.1$--$3.2~\mathrm{GeV}$. 
However, when including the unquenched coupled-channel effects, the bare mass of $D_s(3^1S_0)$ is shifted down to the physical mass
\begin{align}
	M_{\rm phy}\simeq3089~\mathrm{MeV},
\end{align}
which is close to the upper limit of the measured mass of $D_{sJ}(3040)^+$, $M_{\rm exp}=3044\pm8^{+30}_{-5}~\mathrm{MeV}$.
As listed in Table~\ref{tab:Ds_3S_channel_ccef}, the total width of $D_s(3^1S_0)$ is predicted to be $\Gamma \simeq178~\mathrm{MeV}$, and the state mainly decays into the $D^*K$, $D^*_0(2300)K$, $D_s^*\eta$, $D_{2}^*(2460)K$ channels with branching fractions of $\sim 60\%$, $\sim 19\%$, $\sim 8\%$, and $\sim 8\%$, respectively.
If we recalculate the decays with the central value $M_{\rm exp}=3044~\mathrm{MeV}$ for $D_{sJ}(3040)^+$, the total width becomes
\begin{align}
	\Gamma\simeq160~\mathrm{MeV}.
\end{align}
Our predicted width ($\Gamma\simeq 160$--$178~\mathrm{MeV}$) is also compatible with the measured width considering its large uncertainties.
To further confirm the nature of $D_{sJ}(3040)^+$ and finally establish the $D_s(3^1S_0)$, the observation of the $D_s^*\eta$ and $D^*K$ final states and the measurement of the ratio $\Gamma(D_s^*\eta)/\Gamma(D^*K)$ are crucial.

Combining the predicted mass and broad width, the $D_s(3^1S_0)$ assignment is favored over the $D_s(2P_1^\prime)$ one for the $D_{sJ}(3040)^+$.
The broad $D_{sJ}(3040)^+$ structure in the $D^*K$ final state may also receive contributions from several overlapping states, such as $D_s(3^1S_0)$ and $D_s(2P_1^\prime)$.
More experimental observations of the $D_s^*\eta$ and $D_1(2420)K$ final states around the mass range of $3.0$ GeV may be useful to establish these states and clarify the puzzle about the $D_{sJ}(3040)^+$ structure.

\subsection{$D(3000)^0$}

In the $D$-meson family, the $D(3000)^0$ listed in the RPP~\cite{ParticleDataGroup:2024cfk} does not represent a single state from experiments, because the reported structures come from different production processes with different masses and widths.
In 2013, the LHCb collaboration analyzed the $D^+\pi^-$, $D^0\pi^+$, and $D^{*+}\pi^-$ mass spectra~\cite{LHCb:2013jjb}.
In the $D^{*+}\pi^-$ channel, they found a structure $D_J(3000)^0$ with mass $M_{\rm exp}=2971.8\pm8.7~\mathrm{MeV}$ and width $\Gamma_{\rm exp}=188.1\pm44.8~\mathrm{MeV}$; its angular distribution favors an unnatural parity assignment~\cite{LHCb:2013jjb}. 
Meanwhile, in the $D^+\pi^-$ channel, a natural parity structure $D_J^*(3000)^0$ with mass $M_{\rm exp}=3008.1\pm4.0~\mathrm{MeV}$ and width $\Gamma_{\rm exp}=110.5\pm11.5~\mathrm{MeV}$ was observed.
In 2016, by an amplitude analysis of $B^-\to D^+\pi^-\pi^-$, the LHCb collaboration extracted a spin-2 resonance $D_2^*(3000)^0$ with mass $M_{\rm exp}=3214\pm29\pm33\pm36~\mathrm{MeV}$ and width $\Gamma_{\rm exp}=186\pm38\pm34\pm63~\mathrm{MeV}$~\cite{LHCb:2016lxy}.
Although its central parameters differ from those of the previously observed resonance $D_J^*(3000)^0$, one cannot exclude that they
correspond to the same state due to the large uncertainties.

Theoretically, the assignments for these structures are still under debate.
In Ref.~\cite{Wang:2013tka}, the possibilities of the $3S$, $2P$, and $1F$ assignments were discussed with heavy meson effective theory.
Quark model studies suggested various candidates, including the $2P(1^+)$, $2^3P_0$, $3S$, $D(1^3F_4)$, $D(2^3P_2)$, and the higher-mass mixed state $D(2P_1^\prime)$~\cite{Sun:2013qca,Yu:2014dda,Xiao:2014ura,Lu:2014zua,Godfrey:2015dva,Song:2015fha,Li:2017zng,Gupta:2018zlg,Gandhi:2019lta,Ni:2021pce}.
For the spin-2 resonance $D_2^*(3000)^0$, subsequent studies mainly tested the $2P(2^+)$ and $1F(2^+)$ interpretations~\cite{Wang:2016krl,Yu:2016mez,Gupta:2018zlg,Gandhi:2019lta,Zhang:2025kxj}.

Based on our unquenched quark model analysis, we list the predicted masses and widths for the possible candidates $D(3S)$, $D(2P)$, $D(1F)$, and $D(2D)$ that have been proposed in the literature for the $D_J(3000)^0$ and $D_J^*(3000)^0$ signals:
\begin{align}
D(3S):\quad M_{\rm phy}&=2924,\ 3080~\mathrm{MeV}, \nonumber \\
\Gamma&=332,\ 307~\mathrm{MeV},\\
D(2P):\quad M_{\rm phy}&=2673\text{--}2900~\mathrm{MeV},\nonumber \\
\Gamma&=74\text{--}294~\mathrm{MeV},\\
D(1F):\quad M_{\rm phy}&=2965\text{--}3079~\mathrm{MeV}, \nonumber \\
\Gamma&=16\text{--}59~\mathrm{MeV}, \\
D(2D):\quad M_{\rm phy}&=2963\text{--}3176~\mathrm{MeV},\nonumber \\
\Gamma&=89\text{--}131~\mathrm{MeV}.
\end{align}
More details of the decay properties can be seen in Tables~\ref{tab:D_3S_channel_ccef}, \ref{tab:D_2P_channel_ccef},
\ref{tab:D_2D_channel_ccef}, and \ref{tab:D_1F_channel_ccef}.

It is found that although the two $3S$ states $D(3^1S_0)$ and $D(3^3S_1)$ have masses near $3.0~\mathrm{GeV}$, their predicted widths ($\Gamma=332$ and $307~\mathrm{MeV}$) are significantly larger than the measured values $100$--$200~\mathrm{MeV}$ for $D_J^{(*)}(3000)^0$.
For the $D(2P)$ states, although they were proposed as possible candidates in our earlier quenched analyses~\cite{Ni:2021pce}, the coupled-channel effects shift their physical masses down to $2673$--$2900~\mathrm{MeV}$. 
Since these predicted masses lie well below the $3.0~\mathrm{GeV}$ region, the $D(2P)$ interpretations are disfavored by the present unquenched calculation.
For the $D(1F)$ states, although the predicted masses lie in the mass region of $D_J^{(*)}(3000)^0$, the predicted widths ($\Gamma=16$--$59~\mathrm{MeV}$) are too narrow to explain these broad structures.
For the $D(2D)$ states, although the predicted masses and widths are comparable with the observations of $D_J^{(*)}(3000)^0$, their dominant decay modes are inconsistent with the observed $D^{(*)}\pi$.

Therefore, none of the considered $3S$, $2P$, $2D$, or $1F$ states can consistently explain both the masses and decay properties of the $D_J(3000)^0$ and $D_J^*(3000)^0$ structures.
In theory, there indeed exist several states with masses in the range of $\sim3.0$ GeV.
However, the observed decay properties are inconsistent with our predictions based on the unquenched quark model framework.
These observed structures may arise from several highly overlapping states which dominantly decay into $D\pi$ and/or $D^{*}\pi$.
To resolve these puzzles, more precise measurements of the $D^{(*)}\pi$ channels together with more observations of the other channels, such as $D_1(2420)\pi$, $D_2^*(2460)\pi$, and $D\eta$, are needed in future experiments.

\subsection{Missing states and observation strategies}

To facilitate future experimental searches, in Table~\ref{tab:higher_dominant_branching}, we list the dominant decay channels of the missing higher excitations in the $D$- and $D_s$-meson families.
Some states mainly decay into the ground state $D^{(*)}\pi$ and $D^{(*)}K$ channels, while many states have large branching fractions into excited charmed mesons.
Because $D_1(2420)$ and $D_2^*(2460)$ are relatively narrow and can be cleanly reconstructed, the final states containing them are optimal channels for future observations.

If the $D_{sJ}(3040)^+$ indeed corresponds to the $D_s(3^1S_0)$ state, its vector partner $D_s(3^3S_1)$ may also have discovery potential.
For this vector state, the physical mass is predicted to be
\begin{align}
	M_{\rm phy}\simeq3173~\mathrm{MeV},
\end{align}
which lies about $80$ MeV above the $D_s(3^1S_0)$. 
The $D_s(3^3S_1)$ has a relatively broad width of $\Gamma\simeq 150$ MeV, and mainly decays into the $D^*K$ ($\sim 25\%$), $DK$ ($\sim 9\%$), $D^*_1(2600)K$ ($\sim 17\%$), and $D_0(2550)K$ ($\sim 17\%$).
To establish the $D_s(3^3S_1)$ state, the $D^*K$ and $DK$ channels are expected to be observed in future experiments.

If the $D_{s1}(2933)^+$ is indeed assigned as the low mass axial vector state $D_s(2P_1)$, the axial vector states $D(2P_1)$ and $D(2P_1')$ in the charmed sector should also be experimentally accessible.
Considering the coupled-channel effects, the physical masses for the $D(2P_1)$ and $D(2P_1')$ states are predicted to be
\begin{align}
	M_{\rm phy}\simeq2775~\mathrm{MeV}, ~2830 ~\mathrm{MeV},
\end{align}
respectively, which are significantly ($60$--$191~\mathrm{MeV}$) smaller than the masses predicted within the quenched quark models~\cite{Zeng:1994vj,Ebert:2009ua,Li:2010vx,Godfrey:2015dva,Ni:2021pce}.
Both $D(2P_1)$ and $D(2P_1')$ are broad states with a comparable width of $\Gamma\simeq175$ MeV.
From Table~\ref{tab:higher_dominant_branching}, it is found that the $D^*\pi$ and $D_2^*(2460)\pi$ are ideal channels for searching for the two axial vector states in the mass range of $\sim 2.8$ GeV.
Observing these axial vector states, whose masses lie outside the conventional quark model expectations, can help to test the coupled-channel effects.

Additionally, among the $2P$ states, the two tensor states $D(2^3P_2)$ with a mass of $2900$ MeV and $D_s(2^3P_2)$ with a mass of $3039$ MeV have good potential to be observed in experiments.
They are relatively narrow states with widths of $\Gamma\simeq50$--$70~\mathrm{MeV}$.
In contrast, for the two scalar states $D(2^3P_0)$ and $D_s(2^3P_0)$, it may be relatively difficult to establish them in experiments, since they have very broad widths of $\Gamma\simeq200$--$300~\mathrm{MeV}$. 
Due to the strong coupled-channel effects, their masses are shifted to
\begin{align}
	M_{\rm phy}\simeq2673~\mathrm{MeV},~2864 ~\mathrm{MeV},
\end{align}
respectively, which are about $80$--$260~\mathrm{MeV}$ smaller than those predicted within the quenched quark models~\cite{Godfrey:2015dva,Ni:2021pce,Ebert:2009ua}.
The optimal channels for observations are summarized as follows:
\begin{align}
D(2^3P_2)(2900):&\quad D_2^*(2460)\pi\ [\sim35\%], D\pi\ [\sim10\%],\nonumber \\
D_s(2^3P_2)(3039):&\quad D_2^*(2460)K\ [\sim26\%],DK\ [\sim20\%],\nonumber\\
D(2^3P_0)(2673):&\quad D\pi\ [\sim65\%],~D_sK\ [\sim17\%], ~D\eta\ [\sim8\%],\nonumber\\
D_s(2^3P_0)(2864):&\quad DK\ [\sim88\%],~D_s\eta\ [\sim12\%].\nonumber
\end{align}

The $2D$ states in the charmed sector have broad widths of $\Gamma\sim 100$ MeV, while those in the charmed-strange sector are only a few tens of MeV.
To search for the missing $2D$ states, the optimal channels for observations are summarized as follows:
\begin{align}
D(2^3D_1)(2963):&\quad D_1(2420)\pi\ [\sim46\%],\nonumber \\
D(2D_2)(3069):&\quad D_2^*(2460)\pi\ [\sim37\%],\nonumber \\
D(2D_2')(3156):&\quad D_2^*(2460)\pi\ [\sim9\%],\nonumber \\
D_s(2^3D_1)(3165):&\quad D_1(2420)K\ [\sim85\%],\nonumber \\
D_s(2D_2)(3240):&\quad D_2^*(2460)K\ [\sim76\%].\nonumber\\
D_s(2D_2')(3271):&\quad D_2^*(2460)K\ [\sim17\%].\nonumber
\end{align}
More details about the properties of the $2D$ states can be seen in Tables~\ref{tab:D_2D_channel_ccef} and ~\ref{tab:Ds_2D_channel_ccef} in the Appendix.

For the $1F$ states, the predicted total widths are generally narrow, lying in the range of $\Gamma\simeq16$--$59$ MeV.
The optimal channels for observing these $1F$ states are summarized as follows:
\begin{align}
D(1^3F_2)(2990):&\quad D\pi\ [\sim24\%],\ D_1(2420)\pi\ [\sim11\%],\nonumber \\
D(1F_3)(2965):&\quad D_3^*(2750)\pi\ [\sim84\%],\ D_2^*(2460)\pi\ [\sim8\%], \nonumber \\
D(1F'_3)(3079):&\quad D^*\pi\ [\sim27\%],\ D_1(2420)\pi\ [\sim13\%],\nonumber \\
D(1^3F_4)(3014):&\quad D^*\pi\ [\sim33\%],\nonumber\\
D_s(1^3F_2)(3154):&\quad D_1(2420)K\ [\sim57\%],\nonumber \\
D_s(1F_3)(3128):&\quad D_2^*(2460)K\ [\sim82\%],\nonumber \\
D_s(1F'_3)(3176):&\quad D^*K\ [\sim41\%],\ D_1(2420)K\ [\sim14\%],\nonumber \\
D_s(1^3F_4)(3125):&\quad D^*K\ [\sim31\%],\ DK\ [\sim22\%].\nonumber
\end{align}
More details about the properties of the $1F$ states can be seen in Tables \ref{tab:D_1F_channel_ccef} and \ref{tab:Ds_1F_channel_ccef} in the Appendix.

\section{Summary}
\label{sec:summary}

In this work, as a continuation of our previous study on the low-lying heavy-light meson spectra, we systematically study the mass spectra and OZI-allowed strong decays of the higher $3S$-, $2P$-, $2D$-, and $1F$-wave charmed and charmed-strange mesons within an unquenched quark model.
By coupling the bare meson states to the open flavor two meson continua, we systematically calculate the mass shifts, physical masses, and partial decay widths without introducing additional free parameters.

It is found that coupled-channel effects induce sizable state dependent mass shifts, which significantly modify the relative level spacings of the higher excitations. 
These mass shifts depend strongly on the distances to channel thresholds, the decay partial waves, and the radial nodes of the wave functions. 
Meanwhile, the decay widths are determined by the on shell phase space and suppressed by the centrifugal barrier in higher partial waves, and therefore do not simply increase with the number of open channels.

Combining the calculated physical mass, total width, and dominant $D^*K$ decay channel, the newly observed $D_{s1}(2933)^+$ can be assigned as the low mass axial vector state $D_s(2P_1)$.
Its partner $D_s(2P_1^\prime)$ is predicted to lie about $30~\mathrm{MeV}$ higher, which can be searched for in the $D_1(2420)K$ channel to further confirm this assignment.

For the broad structure $D_{sJ}(3040)^+$, our calculated mass, width, and dominant $D^*K$ branching fraction strongly favor the $D_s(3^1S_0)$ assignment, while the alternative $D_s(2P_1^\prime)$ interpretation remains a viable possibility.
To clarify its nature, future experimental searches in the $D_s^*\eta$ and $D_2^*(2460)K$ channels will be crucial.

For the puzzling $D_J^{(*)}(3000)^0$ structures, our results show that none of the conventional $3S$, $2P$, $2D$, or $1F$ states can provide a consistent single state description: the $3S$ states are too broad, the $1F$ states are too narrow, the $D(2P)$ masses lie well below $2.9~\mathrm{GeV}$, and the dominant decay modes of the $2D$ states are inconsistent with the observed $D^{(*)}\pi$ channels.

Finally, it should be emphasized that some missing higher excitations have good potentials to be discovered in future experiments.
Besides the conventional channels $D^{(*)}\pi/K$ and $D_{(s)}\eta$, the $D_1(2420)\pi/K$ and $D_2^*(2460)\pi/K$ channels are expected to be observed as well.

\begin{table*}[!t]
	\caption{Calculated physical masses, total strong widths, and dominant OZI-allowed decay modes for the higher excited $D$ and $D_s$ mesons. Up to three modes are listed for each state, corresponding to the largest nonzero partial widths at the calculated physical mass, normalized to the corresponding total width. The masses and widths are in MeV.}
	\label{tab:higher_dominant_branching}
	\setlength{\tabcolsep}{3.0pt}
	\renewcommand{\arraystretch}{1.06}
 \begin{tabular}{lcccccccccccccccccccccccc}
		\hline\hline
		State & ~~~~~~~~~~~~$M_{\rm phy}$~~~~~~~~~~~~~~ & ~~~~~~~~~~~~~$\Gamma_{\rm total}$~~~~~~~~~~~~~ & ~~~~~~~~~~~~~~~~~~~~~~~~~~~~~~~~~~~~~~~~~Dominant decay modes~~~~~~~~~~~~~~~~~~~~~~~~~~~~~~~~~~~~~~~~~ \\
		\hline
%		\multicolumn{4}{@{}l}{\textit{$D$ mesons}} \\
		$D(3^1S_0)$ & 2924 & 332 & $D^*\pi$ [$\sim 44\%$], $D_1^*(2600)\pi$ [$\sim 20\%$], $D_s^*K$ [$\sim 10\%$] \\
		$D(3^3S_1)$ & 3080 & 307 & $D_1^*(2600)\pi$ [$\sim 28\%$], $D^*\pi$ [$\sim 17\%$], $D_0(2550)\pi$ [$\sim 16\%$] \\
		$D(2^3P_0)$ & 2673 & 294 & $D\pi$ [$\sim 65\%$], $D_sK$ [$\sim 17\%$], $D\eta$ [$\sim 8\%$] \\
		$D(2P_1)$ & 2775 & 174 & $D^*\pi$ [$\sim 37\%$], $D_2^*(2460)\pi$ [$\sim 17\%$], $D_1(2430)\pi$ [$\sim 12\%$] \\
		$D(2P_1^\prime)$ & 2830 & 176 & $D_0^*(2300)\pi$ [$\sim 26\%$], $D_1(2420)\pi$ [$\sim 25\%$], $D^*\pi$ [$\sim 19\%$] \\
		$D(2^3P_2)$ & 2900 & 74 & $D_2^*(2460)\pi$ [$\sim 35\%$], $D_1(2430)\pi$ [$\sim 24\%$], $D\pi$ [$\sim 10\%$] \\
		$D(2^3D_1)$ & 2963 & 128 & $D_1(2420)\pi$ [$\sim 46\%$], $D(1^3D_1)\pi$ [$\sim 27\%$], $D(2P_1)\pi$ [$\sim 12\%$] \\
		$D(2D_2)$ & 3069 & 131 & $D_2^*(2460)\pi$ [$\sim 37\%$], $D_3^*(2750)\pi$ [$\sim 21\%$], $D(2^3P_2)\pi$ [$\sim 18\%$] \\
		$D(2D'_2)$ & 3156 & 113 & $D(1^3D_1)\pi$ [$\sim 43\%$], $D(1D'_2)\pi$ [$\sim 17\%$], $D_2^*(2460)\pi$ [$\sim 9\%$] \\
		$D(2^3D_3)$ & 3176 & 89 & $D_3^*(2750)\pi$ [$\sim 25\%$], $D(1D_2)\pi$ [$\sim 22\%$], $D(2P_1)\pi$ [$\sim 11\%$] \\
		$D(1^3F_2)$ & 2990 & 16 & $D\pi$ [$\sim 24\%$], $D(2P_1)\pi$ [$\sim 12\%$], $D_1(2420)\pi$ [$\sim 11\%$] \\
		$D(1F_3)$ & 2965 & 59 & $D_3^*(2750)\pi$ [$\sim 84\%$], $D_2^*(2460)\pi$ [$\sim 8\%$], $D^*\pi$ [$\sim 4\%$] \\
		$D(1F'_3)$ & 3079 & 54 & $D^*\pi$ [$\sim 27\%$], $D_0^*(2300)\pi$ [$\sim 15\%$], $D_1(2420)\pi$ [$\sim 13\%$] \\
		$D(1^3F_4)$ & 3014 & 33 & $D^*\pi$ [$\sim 33\%$], $D_1(2430)\pi$ [$\sim 27\%$], $D_2^*(2460)\pi$ [$\sim 14\%$] \\
%		\hline\hline
%		\multicolumn{4}{@{}l}{\textit{$D_s$ mesons}} \\
		$D_s(3^1S_0)$ & 3089 & 178 & $D^*K$ [$\sim 60\%$], $D_0^*(2300)K$ [$\sim 19\%$], $D_s^*\eta$ [$\sim 8\%$] \\
		$D_s(3^3S_1)$ & 3173 & 150 & $D^*K$ [$\sim 25\%$], $D_1^*(2600)K$ [$\sim 17\%$], $D_0(2550)K$ [$\sim 17\%$] \\
		$D_s(2^3P_0)$ & 2864 & 201 & $DK$ [$\sim 88\%$], $D_s\eta$ [$\sim 12\%$] \\
		$D_s(2P_1)$ & 2972 & 92 & $D^*K$ [$\sim 59\%$], $DK^*$ [$\sim 18\%$], $D_s^*\eta$ [$\sim 10\%$] \\
		$D_s(2P_1^\prime)$ & 3004 & 96 & $D^*K$ [$\sim 34\%$], $D_1(2420)K$ [$\sim 26\%$], $D_0^*(2300)K$ [$\sim 23\%$] \\
		$D_s(2^3P_2)$ & 3039 & 49 & $D_1(2430)K$ [$\sim 29\%$], $D_2^*(2460)K$ [$\sim 26\%$], $DK$ [$\sim 20\%$] \\
		$D_s(2^3D_1)$ & 3165 & 75 & $D_1(2420)K$ [$\sim 85\%$], $D_2^*(2460)K$ [$\sim 5\%$], $D_1(2430)K$ [$\sim 4\%$] \\
		$D_s(2D_2)$ & 3240 & 59 & $D_2^*(2460)K$ [$\sim 76\%$], $D_{s2}^*(2573)\eta$ [$\sim 5\%$], $DK^*$ [$\sim 5\%$] \\
		$D_s(2D'_2)$ & 3271 & 45 & $D(1^3D_1)K$ [$\sim 57\%$], $D_2^*(2460)K$ [$\sim 17\%$], $D_1^*(2600)K$ [$\sim 9\%$] \\
		$D_s(2^3D_3)$ & 3288 & 31 & $DK$ [$\sim 22\%$], $D^*K$ [$\sim 16\%$], $D_0(2550)K$ [$\sim 13\%$] \\
		$D_s(1^3F_2)$ & 3154 & 23 & $D_1(2420)K$ [$\sim 57\%$], $D_2^*(2460)K$ [$\sim 11\%$], $D^*K^*$ [$\sim 10\%$] \\
		$D_s(1F_3)$ & 3128 & 22 & $D_2^*(2460)K$ [$\sim 82\%$], $D^*K$ [$\sim 6\%$], $DK^*$ [$\sim 5\%$] \\
		$D_s(1F'_3)$ & 3176 & 35 & $D^*K$ [$\sim 41\%$], $D_0^*(2300)K$ [$\sim 19\%$], $D_1(2420)K$ [$\sim 14\%$] \\
		$D_s(1^3F_4)$ & 3125 & 21 & $D^*K$ [$\sim 31\%$], $D_1(2430)K$ [$\sim 26\%$], $DK$ [$\sim 22\%$] \\
		\hline\hline
	\end{tabular}
\end{table*}

\begin{acknowledgments}
This work is supported by the National Natural Science Foundation of China (Grants No. 12221005, No. 12235018, No. 12175065), and the Chinese Academy of Sciences under Grant No. YSBR-101, and the National Key Research and Development Program of China under Contract No. 2025YFA1613900.
\end{acknowledgments}

\section*{DATA AVAILABILITY}
All data supporting the findings of this article are available within the article.

\appendix*

\section{Channel contributions to mass shifts and strong decay widths}
\label{app:channel_decomposition}

The channel contributions to the mass shifts and OZI-allowed strong decay widths of the higher excited $D$ and $D_s$ mesons are listed in Tables~\ref{tab:D_3S_channel_ccef}--\ref{tab:Ds_1F_channel_ccef}.
These tables are arranged in the order of the $3S$-, $2P$-, $2D$-, and $1F$-wave states for the $D$ and $D_s$ mesons, respectively.

\begin{table}[!t]
\caption{Coupled-channel contributions to the mass shifts $\Delta M_i$, and OZI-allowed partial widths $\Gamma_i$ for the $D(3S)$ states. The numbers in square brackets after $\Delta M_i$ and $\Gamma_i$ are the bare mass $M_A^{(0)}$ and the physical mass $M_{\rm phy}$, respectively. For unestablished final heavy mesons, the parenthesized numbers denote the masses adopted in the calculation. An ellipsis denotes a channel not included for the corresponding initial state. All entries are in MeV.}
\label{tab:D_3S_channel_ccef}
\small
\setlength{\tabcolsep}{1.80pt}
\renewcommand{\arraystretch}{1.05}
\begin{tabular}{ccccccccccccccccccccccccccccccccccccccc}
\hline\hline
 & \multicolumn{2}{c}{\rule[-0.5ex]{0pt}{2.6ex}$D(3^1S_0)$} && \multicolumn{2}{c}{\rule[-0.5ex]{0pt}{2.6ex}$D(3^3S_1)$} \\
\cline{2-3}\cline{5-6}
Channel & $\Delta M_i [3071]$ & $\Gamma_i [2924]$ && $\Delta M_i [3136]$ & $\Gamma_i [3080]$ \\
\hline
$D \pi$ & $\cdots$ & $\cdots$ && 11.6 & 27.2 \\
$D_s K$ & $\cdots$ & $\cdots$ && 1.1 & 11.1 \\
$D \eta$ & $\cdots$ & $\cdots$ && 0.8 & 4.2 \\
$D \eta'$ & $\cdots$ & $\cdots$ && -0.4 & 2.2 \\
$D^* \pi$ & -15.4 & 146.5 &&8.5 & 52.8 \\
$D_s^* K$ & -12.7 & 34.2 && -2.1 & 20.3 \\
$D^* \eta$ & -4.7 & 15.7 && -0.3 & 8.1 \\
$D^* \eta'$ & -1.9 & 0.0 && -1.2 & 1.6 \\
$D_0(2550) \pi$ & $\cdots$ & $\cdots$ && -0.8 & 48.1 \\
$D_{s0}(2590) K$ & $\cdots$ & $\cdots$ && -1.5 & 0.0 \\
$D_0(2550) \eta$ & $\cdots$ & $\cdots$ && -0.7 & 0.0 \\
$D_1^*(2600) \pi$ & -47.3 & 67.5 && -14.0 & 87.3 \\
$D_0^*(2300) \pi$ & -17.0 & 24.0 && $\cdots$ & $\cdots$ \\
$D_{s0}^*(2317) K$ & -11.5 & 17.7 && $\cdots$ & $\cdots$ \\
$D_0^*(2300) \eta$ & -3.0 & 2.4 && $\cdots$ & $\cdots$ \\
$D_1(2430) \pi$ & $\cdots$ & $\cdots$ && -12.6 & 16.3 \\
$D_{s1}(2460) K$ & $\cdots$ & $\cdots$ && -7.5 & 9.2 \\
$D_1(2430) \eta$ & $\cdots$ & $\cdots$ && -2.4 & 2.8 \\
$D_1(2420) \pi$ & $\cdots$ & $\cdots$ && -6.8 & 0.5 \\
$D_{s1}(2536) K$ & $\cdots$ & $\cdots$ && -2.0 & $4.8 \times 10^{-2}$ \\
$D_1(2420) \eta$ & $\cdots$ & $\cdots$ && -0.9 & 0.2 \\
$D_2^*(2460) \pi$ & -11.7 & 17.0 && -4.5 & 5.2 \\
$D_{s2}^*(2573) K$ & -2.3 & 0.0 && -1.6 & $7.5 \times 10^{-3}$ \\
$D_2^*(2460) \eta$ & -1.2 & 0.0 && -0.8 & 0.2 \\
$D(1^3D_1)(2648) \pi$ & -3.1 & 5.5 && -1.3 & 6.2 \\
$D(1D_2)(2723) \pi$ & $\cdots$ & $\cdots$ && -1.7 & 0.5 \\
$D(1D'_2)(2789) \pi$ & $\cdots$ & $\cdots$ && -5.7 & $3.6 \times 10^{-2}$ \\
$D_3^*(2750) \pi$ & -14.9 & $3.5 \times 10^{-4}$ && -7.3 & 0.5 \\
$D \rho$ & $-3.0 \times 10^{-2}$ & $1.6 \times 10^{-2}$ && -0.5 & 1.8 \\
$D \omega$ & $-1.0 \times 10^{-2}$ & $4.5 \times 10^{-3}$ && -0.2 & 0.6 \\
$D_s K^*$ & $2.5 \times 10^{-3}$ & $4.0 \times 10^{-2}$ && -0.2 & 0.3 \\
$D^* \rho$ & -0.3 & 0.9 && -0.4 & 0.1 \\
$D^* \omega$ & -0.1 & 0.3 && -0.1 & $1.8 \times 10^{-2}$ \\
$D_s^* K^*$ & -0.1 & 0.0 && -0.2 & $2.5 \times 10^{-2}$ \\
$D_0^*(2300) \rho$ & $\cdots$ & $\cdots$ && -0.5 & 0.0 \\
$D_0^*(2300) \omega$ & $\cdots$ & $\cdots$ && -0.2 & 0.0 \\
\hline
$\mathrm{Total}$ & -147.2 & 331.8 && -56.4 & 307.4 \\
\hline\hline
\end{tabular}
\end{table}

\begin{table*}[!t]
\caption{Channel contributions to the mass shifts $\Delta M_i$ and OZI-allowed partial decay widths $\Gamma_i$ for the $D_s(3S)$ states. The notation and units are the same as in Table~\ref{tab:D_3S_channel_ccef}. }
\label{tab:Ds_3S_channel_ccef}
\small
\setlength{\tabcolsep}{15.00pt}
\renewcommand{\arraystretch}{1.05}
\begin{tabular}{ccccccccccccccc}
\hline\hline
 & \multicolumn{2}{c}{\rule[-0.5ex]{0pt}{2.6ex}$D_s(3^1S_0)$} & & \multicolumn{2}{c}{\rule[-0.5ex]{0pt}{2.6ex}$D_s(3^3S_1)$} \\
\cline{2-3}\cline{5-6}
Channel & $\Delta M_i [3160]$ & $\Gamma_i [3089/\mathrm{as}~D_{sJ}(3040)]$ & & $\Delta M_i [3213]$ & $\Gamma_i [3173]$ \\
\hline
$D K$ & $\cdots$ & $\cdots$ && 5.4 & 14.0 \\
$D_s \eta$ & $\cdots$ & $\cdots$ && 0.4 & 2.4 \\
$D_s \eta'$ & $\cdots$ & $\cdots$ &&-0.5 & 2.9 \\
$D^* K$ & -8.9 & $106.5/101.4$ && 3.3 & 36.8 \\
$D_s^* \eta$ & -2.8 & $14.1/12.5$ && -0.3 & 5.8 \\
$D_s^* \eta'$ & -4.1 & $0.4/0.0$ && -1.9 & 2.3 \\
$D_0(2550) K$ & $\cdots$ & $\cdots$ && -4.8 & 25.5 \\
$D_{s0}(2590) \eta$ & $\cdots$ & $\cdots$ && -0.7 & 0.8 \\
$D_1^*(2600) K$ & -23.7 & $0.0/0.0$ && -17.6 & 25.9 \\
$D_0^*(2300) K$ & -13.0 & $33.3/30.9$ && $\cdots$ & $\cdots$ \\
$D_{s0}^*(2317) \eta$ & -3.0 & $8.7/8.0$ && $\cdots$ & $\cdots$ \\
$D_1(2430) K$ & $\cdots$ & $\cdots$ && -8.5 & 17.6 \\
$D_{s1}(2460) \eta$ & $\cdots$ & $\cdots$ && -1.8 & 3.1 \\
$D_1(2420) K$ & $\cdots$ & $\cdots$ && -5.3 & 2.7 \\
$D_{s1}(2536) \eta$ & $\cdots$ & $\cdots$ && -0.7 & 0.1 \\
$D_2^*(2460) K$ & -12.9 & $13.9/6.1$ && -4.9 & 8.1 \\
$D_{s2}^*(2573) \eta$ & -1.1 & $0.0/0.0$ && -0.6 & 0.1 \\
$D(1^3D_1)(2648) K$ & -1.3 & $0.0/0.0$ && -0.4 & $4.9 \times 10^{-2}$ \\
$D K^*$ & $-4.7 \times 10^{-2}$ & $0.1/0.1$ && -0.6 & 1.7 \\
$D_s \phi$ & $-1.9 \times 10^{-2}$ & $7.5\times 10^{-4}/2.3\times 10^{-6}$ && -0.2 & 0.2 \\
$D^* K^*$ & -0.3 & $1.3/0.9$ && -0.4 & $9.3 \times 10^{-4}$ \\
$D_s^* \phi$ & -0.1 & $0.0/0.0$ && -0.1 & $8.4 \times 10^{-4}$ \\
\hline
$\mathrm{Total}$ & -71.3 & $178.3/159.9$ && -40.2 & 150.1 \\
\hline\hline
\end{tabular}
\end{table*}

\begin{table*}[t!]
\caption{Channel contributions to the mass shifts $\Delta M_i$ and OZI-allowed partial widths $\Gamma_i$ for the $D(2P)$ states. The notation and units are the same as in Table~\ref{tab:D_3S_channel_ccef}.}
\label{tab:D_2P_channel_ccef}
\small
\setlength{\tabcolsep}{1.25pt}
\renewcommand{\arraystretch}{1.05}
\begin{tabular}{ccccccccccccccccccccccccccc}
\hline\hline
 & \multicolumn{2}{c}{\rule[-0.5ex]{0pt}{2.6ex}$\underline{~~~~~~~~~~~~~~D(2^3P_0)~~~~~~~~~~~~~~}$} & \multicolumn{2}{c}{\rule[-0.5ex]{0pt}{2.6ex}$\underline{~~~~~~~~~~~~D(2P_1)~~~~~~~~~~~~}$} & \multicolumn{2}{c}{\rule[-0.5ex]{0pt}{2.6ex}$\underline{~~~~~~~~~~~~D(2P_1^\prime)~~~~~~~~~~~~}$} & \multicolumn{2}{c}{\rule[-0.5ex]{0pt}{2.6ex}$\underline{~~~~~~~~~~~~D(2^3P_2)~~~~~~~~~~~~}$} \\
%\cline{2-3}\cline{4-5}\cline{6-7}\cline{8-9}
~~~~Channel~~~~ & ~~~~$\Delta M_i [2871]$ ~~&~~ $\Gamma_i [2673]$~~~~ & ~~~~$\Delta M_i [2937]$ ~~& ~~$\Gamma_i [2775]$ ~~~~& ~~~~$\Delta M_i [2970]$ ~~& ~~$\Gamma_i [2830]$ ~~~~& ~~~~$\Delta M_i [3002]$ ~~& ~~$\Gamma_i [2900]$ \\
\hline
$D \pi$ & -54.5 & 189.9 & $\cdots$ & $\cdots$ & $\cdots$ & $\cdots$ & -6.5 & 7.5 \\
$D_s K$ & -29.4 & 48.5 & $\cdots$ & $\cdots$ & $\cdots$ & $\cdots$ & -1.7 & 1.0 \\
$D \eta$ & -12.0 & 24.5 & $\cdots$ & $\cdots$ & $\cdots$ & $\cdots$ & -0.9 & 0.6 \\
$D \eta'$ & -4.5 & 0.0 & $\cdots$ & $\cdots$ & $\cdots$ & $\cdots$ & -0.5 & $7.2 \times 10^{-4}$ \\
$D^* \pi$ & $\cdots$ & $\cdots$ & -18.9 & 64.6 & -7.1 & 33.7 & -5.5 & 1.1 \\
$D_s^* K$ & $\cdots$ & $\cdots$ & -13.0 & 20.2 & -7.0 & 13.0 & -1.6 & $4.2 \times 10^{-3}$ \\
$D^* \eta$ & $\cdots$ & $\cdots$ & -4.5 & 8.6 & -2.3 & 5.4 & -0.7 & $3.2 \times 10^{-5}$ \\
$D^* \eta'$ & $\cdots$ & $\cdots$ & $\cdots$ & $\cdots$ & -1.0 & 0.0 & -0.5 & 0.0 \\
$D_0(2550) \pi$ & -58.2 & 0.0 & $\cdots$ & $\cdots$ & $\cdots$ & $\cdots$ & -5.2 & 3.1 \\
$D_1^*(2600) \pi$ & $\cdots$ & $\cdots$ & -44.7 & 7.2 & -30.7 & 14.0 & -7.3 & 1.3 \\
$D_0^*(2300) \pi$ & $\cdots$ & $\cdots$ & -2.0 & 2.0 & -7.9 & 45.2 & $\cdots$ & $\cdots$ \\
$D_{s0}^*(2317) K$ & $\cdots$ & $\cdots$ & -0.6 & 0.0 & -2.2 & 0.5 & $\cdots$ & $\cdots$ \\
$D_0^*(2300) \eta$ & $\cdots$ & $\cdots$ & -0.3 & 0.0 & -0.6 & 0.0 & $\cdots$ & $\cdots$ \\
$D_1(2430) \pi$ & -6.0 & 7.6 & -5.3 & 20.7 & -2.4 & 0.1 & -6.9 & 17.6 \\
$D_{s1}(2460) K$ & $\cdots$ & $\cdots$ & $\cdots$ & $\cdots$ & -0.7 & 0.0 & -1.6 & 0.0 \\
$D_1(2430) \eta$ & $\cdots$ & $\cdots$ & $\cdots$ & $\cdots$ & -0.3 & 0.0 & -0.7 & 0.0 \\
$D_1(2420) \pi$ & -32.8 & 23.0 & -0.8 & 1.2 & -10.6 & 44.2 & -5.2 & 5.7 \\
$D_1(2420) \eta$ & $\cdots$ & $\cdots$ & $\cdots$ & $\cdots$ & $\cdots$ & $\cdots$ & -0.5 & 0.0 \\
$D_2^*(2460) \pi$ & $\cdots$ & $\cdots$ & -23.1 & 29.7 & -11.0 & 4.6 & -8.1 & 25.8 \\
$D(1^3D_1)(2648) \pi$ & $\cdots$ & $\cdots$ & -19.3 & 0.0 & -41.3 & 10.8 & -1.1 & 0.1 \\
$D(1D_2)(2723) \pi$ & $\cdots$ & $\cdots$ & -0.6 & 0.0 & -0.8 & 0.0 & -26.8 & 0.0 \\
$D(1D'_2)(2789) \pi$ & $\cdots$ & $\cdots$ & $\cdots$ & $\cdots$ & $\cdots$ & $\cdots$ & -7.6 & 0.0 \\
$D_3^*(2750) \pi$ & $\cdots$ & $\cdots$ & -23.6 & 0.0 & -12.0 & 0.0 & -10.0 & $2.4 \times 10^{-5}$ \\
$D \rho$ & $\cdots$ & $\cdots$ & -2.5 & 15.1 & $2.0 \times 10^{-3}$ & 0.9 & -0.5 & 0.6 \\
$D \omega$ & $\cdots$ & $\cdots$ & -0.9 & 4.8 & $-3.8 \times 10^{-3}$ & 0.3 & -0.2 & 0.2 \\
$D_s K^*$ & $\cdots$ & $\cdots$ & -1.2 & 0.0 & -0.1 & 0.0 & -0.1 & $3.2 \times 10^{-3}$ \\
$D^* \rho$ & -0.5 & 0.0 & -0.8 & 0.0 & -1.6 & 2.2 & -1.3 & 6.8 \\
$D^* \omega$ & -0.2 & 0.0 & -0.3 & 0.0 & -0.5 & 0.6 & -0.5 & 2.2 \\
\hline
$\mathrm{Total}$ & -198.1 & 293.5 & -162.4 & 174.1 & -140.1 & 175.5 & -101.5 & 73.6 \\
\hline\hline
\end{tabular}
\end{table*}

\begin{table*}[h!]
\caption{Channel contributions to the mass shifts $\Delta M_i$ and OZI-allowed partial widths $\Gamma_i$ for the $D_s(2P)$ states. The notation and units are the same as in Table~\ref{tab:D_3S_channel_ccef}.}
\label{tab:Ds_2P_channel_ccef}
\small
\setlength{\tabcolsep}{1.25pt}
\renewcommand{\arraystretch}{1.05}
\begin{tabular}{cccccccccccccccccccccccccccccc}
\hline\hline
 & \multicolumn{2}{c}{\rule[-0.5ex]{0pt}{2.6ex}$\underline{~~~~~~~~~~~~~~D_s(2^3P_0)~~~~~~~~~~~~~~}$} & \multicolumn{2}{c}{\rule[-0.5ex]{0pt}{2.6ex}$\underline{~~~~~~~~~~~~~~D_s(2P_1)~~~~~~~~~~~~~~}$} & \multicolumn{2}{c}{\rule[-0.5ex]{0pt}{2.6ex}$\underline{~~~~~~~~~~~~~~D_s(2P_1^\prime)~~~~~~~~~~~~~~}$} & \multicolumn{2}{c}{\rule[-0.5ex]{0pt}{2.6ex}$\underline{~~~~~~~~~~~~~~D_s(2^3P_2)~~~~~~~~~~~~~~}$} \\
%\cline{2-3}\cline{4-5}\cline{6-7}\cline{8-9}
Channel & ~~$\Delta M_i [2944]$ ~~& ~~$\Gamma_i [2864]$ ~~~~&~~~~ $\Delta M_i [3019]$ ~~&~~ $\Gamma_i [2972]$ ~~~~&~~~~ $\Delta M_i [3044]$ ~~&~~ $\Gamma_i [3004]$ ~~~~&~~~~ $\Delta M_i [3083]$ ~~&~~ $\Gamma_i [3039]$ \\
\hline
$D K$ & -32.7 & 177.0 & $\cdots$ & $\cdots$ & $\cdots$ & $\cdots$ & -6.3 & 9.7 \\
$D_s \eta$ & -7.8 & 24.0 & $\cdots$ & $\cdots$ & $\cdots$ & $\cdots$ & -0.8 & 0.9 \\
$D_s \eta'$ & -10.3 & 0.0 & $\cdots$ & $\cdots$ & $\cdots$ & $\cdots$ & -0.9 & $3.0 \times 10^{-2}$ \\
$D^* K$ & $\cdots$ & $\cdots$ & -8.3 & 53.8 & -3.7 & 32.5 & -5.5 & 1.6 \\
$D_s^* \eta$ & $\cdots$ & $\cdots$ & -3.0 & 9.5 & -1.8 & 6.2 & -0.7 & 0.1 \\
$D_s^* \eta'$ & $\cdots$ & $\cdots$ & $\cdots$ & $\cdots$ & $\cdots$ & $\cdots$ & -0.9 & 0.0 \\
$D_0(2550) K$ & $\cdots$ & $\cdots$ & $\cdots$ & $\cdots$ & $\cdots$ & $\cdots$ & -4.9 & 0.0 \\
$D_0^*(2300) K$ & $\cdots$ & $\cdots$ & -2.0 & 0.1 & -4.7 & 21.6 & $\cdots$ & $\cdots$ \\
$D_{s0}^*(2317) \eta$ & $\cdots$ & $\cdots$ & -0.2 & $1.7 \times 10^{-2}$ & -0.6 & 2.2 & $\cdots$ & $\cdots$ \\
$D_1(2430) K$ & -3.3 & 0.0 & -3.5 & 5.3 & -2.0 & 0.8 & -5.1 & 14.4 \\
$D_{s1}(2460) \eta$ & $\cdots$ & $\cdots$ & -0.3 & 0.0 & -0.3 & 0.0 & -0.7 & 0.4 \\
$D_1(2420) K$ & -25.3 & 0.0 & -0.5 & 0.3 & -11.8 & 25.0 & -5.3 & 2.3 \\
$D_{s1}(2536) \eta$ & $\cdots$ & $\cdots$ & $\cdots$ & $\cdots$ & $\cdots$ & $\cdots$ & -0.4 & 0.0 \\
$D_2^*(2460) K$ & $\cdots$ & $\cdots$ & -24.8 & 5.9 & -12.9 & 4.3 & -10.0 & 12.7 \\
$D K^*$ & $\cdots$ & $\cdots$ & -1.6 & 16.8 & -0.1 & 0.3 & -0.6 & 0.9 \\
$D_s \phi$ & $\cdots$ & $\cdots$ & -1.4 & 0.0 & -0.1 & 0.2 & -0.1 & $5.4 \times 10^{-3}$ \\
$D^* K^*$ & -0.8 & 0.0 & -0.9 & 0.1 & -1.5 & 2.5 & -1.3 & 6.4 \\
\hline
$\mathrm{Total}$ & -80.2 & 201.0 & -46.5 & 91.8 & -39.5 & 95.6 & -43.5 & 49.4 \\
\hline\hline
\end{tabular}
\end{table*}

\begin{table*}[h!]
\caption{Coupled-channel contributions to the mass shifts $\Delta M_i$, and OZI-allowed partial widths $\Gamma_i$ for the $D(2D)$ states. The notation and units are the same as in Table~\ref{tab:D_3S_channel_ccef}.}
\label{tab:D_2D_channel_ccef}
\small
\setlength{\tabcolsep}{1.10pt}
\renewcommand{\arraystretch}{1.05}
\begin{tabular}{ccccccccccccccccccccccccccccccccccccccc}
\hline\hline
 & \multicolumn{2}{c}{\rule[-0.5ex]{0pt}{2.6ex}$\underline{~~~~~~~~~~~~~~D(2^3D_1)~~~~~~~~~~~~~~}$} & \multicolumn{2}{c}{\rule[-0.5ex]{0pt}{2.6ex}$\underline{~~~~~~~~~~~~~~D(2D_2)~~~~~~~~~~~~~~}$} & \multicolumn{2}{c}{\rule[-0.5ex]{0pt}{2.6ex}$\underline{~~~~~~~~~~~~~~D(2D'_2)~~~~~~~~~~~~~~}$} & \multicolumn{2}{c}{\rule[-0.5ex]{0pt}{2.6ex}$\underline{~~~~~~~~~~~~~~D(2^3D_3)~~~~~~~~~~~~~~}$} \\
%\cline{2-3}\cline{4-5}\cline{6-7}\cline{8-9}
Channel & $\Delta M_i [3168]$ ~~&~~ $\Gamma_i [2963]$~~~~ &~~~~ $\Delta M_i [3217]$~~ &~~ $\Gamma_i [3069]$ ~~~~&~~~~ $\Delta M_i [3260]$ ~~&~~ $\Gamma_i [3156]$ ~~~~&~~~~ $\Delta M_i [3257]$ & $\Gamma_i [3176]$ \\
\hline
$D \pi$ & -3.4 & 2.4 & $\cdots$ & $\cdots$ & $\cdots$ & $\cdots$ & -3.0 & 6.6 \\
$D_s K$ & -1.1 & 0.6 & $\cdots$ & $\cdots$ & $\cdots$ & $\cdots$ & -0.8 & 1.2 \\
$D \eta$ & -0.5 & 0.2 & $\cdots$ & $\cdots$ & $\cdots$ & $\cdots$ & -0.4 & 0.7 \\
$D \eta'$ & -0.2 & 0.1 & $\cdots$ & $\cdots$ & $\cdots$ & $\cdots$ & -0.3 & 0.1 \\
$D^* \pi$ & -1.7 & $8.1 \times 10^{-5}$ & -4.0 & 0.1 & -4.5 & 2.6 & -4.2 & 5.1 \\
$D_s^* K$ & -0.4 & $4.1 \times 10^{-2}$ & -2.1 & 0.2 & -1.1 & 0.2 & -1.0 & 0.6 \\
$D^* \eta$ & -0.2 & $2.5 \times 10^{-6}$ & -0.8 & 0.1 & -0.7 & 0.1 & -0.5 & 0.4 \\
$D^* \eta'$ & -0.1 & 0.0 & -0.4 & $4.2 \times 10^{-2}$ & -0.4 & $2.1 \times 10^{-2}$ & -0.3 & $9.0 \times 10^{-4}$ \\
$D_0(2550) \pi$ & -0.9 & 7.6 & $\cdots$ & $\cdots$ & $\cdots$ & $\cdots$ & -2.0 & 5.0 \\
$D_{s0}(2590) K$ & -0.4 & 0.0 & $\cdots$ & $\cdots$ & $\cdots$ & $\cdots$ & -0.4 & 0.1 \\
$D_0(2550) \eta$ & -0.2 & 0.0 & $\cdots$ & $\cdots$ & $\cdots$ & $\cdots$ & -0.2 & $2.7 \times 10^{-2}$ \\
$D_1^*(2600) \pi$ & -0.8 & 0.7 & -4.4 & 3.9 & -8.7 & 9.6 & -4.0 & 5.8 \\
$D_{s1}^*(2700) K$ & $\cdots$ & $\cdots$ & -1.3 & 0.0 & -1.5 & 0.0 & -0.6 & 0.0 \\
$D_1^*(2600) \eta$ & $\cdots$ & $\cdots$ & -0.6 & 0.0 & -0.7 & 0.0 & -0.3 & $9.0 \times 10^{-9}$ \\
$D_0^*(2300) \pi$ & $\cdots$ & $\cdots$ & -0.4 & 3.5 & 0.3 & 0.1 & $\cdots$ & $\cdots$ \\
$D_{s0}^*(2317) K$ & $\cdots$ & $\cdots$ & -0.1 & 0.3 & -0.1 & $4.4 \times 10^{-3}$ & $\cdots$ & $\cdots$ \\
$D_0^*(2300) \eta$ & $\cdots$ & $\cdots$ & -0.1 & 0.1 & $-4.0 \times 10^{-2}$ & $4.5 \times 10^{-2}$ & $\cdots$ & $\cdots$ \\
$D_1(2430) \pi$ & -1.5 & 5.2 & -0.6 & 4.4 & 0.1 & 0.1 & 0.3 & $2.1 \times 10^{-2}$ \\
$D_{s1}(2460) K$ & -0.2 & $1.1 \times 10^{-2}$ & -0.2 & 0.1 & -0.1 & 0.1 & -0.2 & 0.1 \\
$D_1(2430) \eta$ & -0.1 & $4.6 \times 10^{-3}$ & -0.1 & $4.6 \times 10^{-2}$ & -0.1 & 0.1 & -0.1 & 0.1 \\
$D_1(2420) \pi$ & -57.2 & 59.2 & -0.7 & $1.1 \times 10^{-2}$ & -1.3 & 3.6 & -2.4 & 2.4 \\
$D_{s1}(2536) K$ & -11.4 & 0.0 & -0.2 & $3.1 \times 10^{-4}$ & -1.0 & 0.3 & -0.5 & $3.2 \times 10^{-2}$ \\
$D_1(2420) \eta$ & -4.9 & 0.0 & -0.1 & $3.0 \times 10^{-3}$ & -0.4 & 0.4 & -0.3 & 0.1 \\
$D_2^*(2460) \pi$ & -1.9 & 2.2 & -26.3 & 48.7 & -5.9 & 9.6 & -3.3 & 3.9 \\
$D_{s2}^*(2573) K$ & -0.3 & 0.0 & -8.1 & 0.9 & -1.3 & 0.3 & -1.3 & 0.2 \\
$D_2^*(2460) \eta$ & -0.2 & 0.0 & -3.5 & 1.4 & -0.7 & 0.3 & -0.6 & 0.3 \\
$D(2^3P_0)(2673) \pi$ & $\cdots$ & $\cdots$ & -1.2 & 0.5 & -2.6 & 8.8 & $\cdots$ & $\cdots$ \\
$D(2^3P_0)(2673) \eta$ & $\cdots$ & $\cdots$ & $\cdots$ & $\cdots$ & -0.3 & 0.0 & $\cdots$ & $\cdots$ \\
$D(2P_1)(2775) \pi$ & -33.9 & 14.7 & -0.3 & $7.1 \times 10^{-3}$ & -0.3 & 0.1 & -8.1 & 9.8 \\
$D(2P_1^\prime)(2830) \pi$ & -42.5 & 0.0 & -1.2 & 0.1 & -5.1 & 3.4 & -4.6 & 0.6 \\
$D(2^3P_2)(2901) \pi$ & -1.1 & 0.0 & -59.5 & 24.0 & -11.5 & 3.0 & -6.8 & 0.8 \\
$D(1^3D_1)(2648) \pi$ & -15.2 & 34.4 & -1.2 & 1.7 & -27.3 & 48.6 & -0.3 & 0.4 \\
$D(1^3D_1)(2648) \eta$ & $\cdots$ & $\cdots$ & -0.2 & 0.0 & -2.8 & 0.0 & $-2.3 \times 10^{-2}$ & 0.0 \\
$D(1D_2)(2723) \pi$ & -0.5 & $6.5 \times 10^{-3}$ & -5.1 & 9.1 & -1.8 & 0.1 & -19.7 & 19.9 \\
$D(1D'_2)(2789) \pi$ & -19.5 & 0.0 & -0.9 & $3.7 \times 10^{-2}$ & -11.2 & 18.8 & -3.8 & 0.1 \\
$D_3^*(2750) \pi$ & -2.9 & $1.0 \times 10^{-3}$ & -20.0 & 27.1 & -9.1 & 0.4 & -7.7 & 21.9 \\
$D \rho$ & $-2.3 \times 10^{-2}$ & $1.3 \times 10^{-3}$ & -0.4 & 2.3 & 0.1 & 0.4 & -0.1 & 0.3 \\
$D \omega$ & $-7.3 \times 10^{-3}$ & $3.2 \times 10^{-4}$ & -0.1 & 0.8 & $2.0 \times 10^{-2}$ & 0.1 & $-3.9 \times 10^{-2}$ & 0.1 \\
$D_s K^*$ & $-6.0 \times 10^{-3}$ & $5.3 \times 10^{-7}$ & -0.3 & 0.6 & $-1.7 \times 10^{-2}$ & 0.3 & $-3.6 \times 10^{-2}$ & $3.2 \times 10^{-2}$ \\
$D^* \rho$ & -0.3 & 0.1 & -0.6 & 0.8 & -0.3 & 0.9 & -0.3 & 1.4 \\
$D^* \omega$ & -0.1 & $1.8 \times 10^{-2}$ & -0.2 & 0.2 & -0.1 & 0.3 & -0.1 & 0.5 \\
$D_s^* K^*$ & -0.1 & 0.0 & -0.2 & $4.5 \times 10^{-2}$ & -0.1 & 0.2 & -0.2 & 0.5 \\
$D_0^*(2300) \rho$ & -1.3 & 0.0 & $-2.5 \times 10^{-2}$ & 0.0 & -0.3 & $7.9 \times 10^{-3}$ & $-1.1 \times 10^{-2}$ & $2.2 \times 10^{-3}$ \\
$D_0^*(2300) \omega$ & -0.4 & 0.0 & $-8.2 \times 10^{-3}$ & 0.0 & -0.1 & $9.5 \times 10^{-4}$ & $-3.6 \times 10^{-3}$ & $4.5 \times 10^{-4}$ \\
$D_{s0}^*(2317) K^*$ & $\cdots$ & $\cdots$ & $-9.8 \times 10^{-3}$ & 0.0 & -0.1 & 0.0 & $-3.0 \times 10^{-3}$ & 0.0 \\
$D_1(2430) \rho$ & $\cdots$ & $\cdots$ & -1.8 & 0.0 & -0.2 & 0.0 & -0.5 & 0.0 \\
$D_1(2430) \omega$ & $\cdots$ & $\cdots$ & -0.6 & 0.0 & -0.1 & 0.0 & -0.1 & 0.0 \\
$D_1(2420) \rho$ & $\cdots$ & $\cdots$ & -0.3 & 0.0 & -1.7 & 0.0 & -0.2 & 0.0 \\
$D_1(2420) \omega$ & $\cdots$ & $\cdots$ & -0.1 & 0.0 & -0.5 & 0.0 & -0.1 & 0.0 \\
$D_2^*(2460) \rho$ & $\cdots$ & $\cdots$ & $\cdots$ & $\cdots$ & -0.3 & 0.0 & -1.5 & 0.0 \\
$D_2^*(2460) \omega$ & $\cdots$ & $\cdots$ & $\cdots$ & $\cdots$ & -0.1 & 0.0 & -0.5 & 0.0 \\
\hline
$\mathrm{Total}$ & -205.4 & 127.5 & -148.2 & 131.1 & -103.9 & 112.9 & -81.1 & 89.1 \\
\hline\hline
\end{tabular}
\end{table*}

\begin{table*}[t]
\caption{Coupled-channel contributions to the mass shifts $\Delta M_i$, and OZI-allowed partial widths $\Gamma_i$ for the $D_s(2D)$ states. The notation and units are the same as in Table~\ref{tab:D_3S_channel_ccef}.}
\label{tab:Ds_2D_channel_ccef}
\small
\setlength{\tabcolsep}{1.25pt}
\renewcommand{\arraystretch}{1.05}
\begin{tabular}{ccccccccccccccccccccccccccccc}
\hline\hline
 & \multicolumn{2}{c}{\rule[-0.5ex]{0pt}{2.6ex}\shortstack{$\underline{~~~~~~~~~~~~~~D_s(2^3D_1)~~~~~~~~~~~~~~}$}} & \multicolumn{2}{c}{\rule[-0.5ex]{0pt}{2.6ex}\shortstack{$\underline{~~~~~~~~~~~~~~D_s(2D_2)~~~~~~~~~~~~~~}$}} & \multicolumn{2}{c}{\rule[-0.5ex]{0pt}{2.6ex}\shortstack{$\underline{~~~~~~~~~~~~~~D_s(2D'_2)~~~~~~~~~~~~~~}$}} & \multicolumn{2}{c}{\rule[-0.5ex]{0pt}{2.6ex}\shortstack{$\underline{~~~~~~~~~~~~~~D_s(2^3D_3)~~~~~~~~~~~~~~}$}} \\
%\cline{2-3}\cline{4-5}\cline{6-7}\cline{8-9}
Channel & ~~$\Delta M_i [3234]$~~ &~~ $\Gamma_i [3165]$ ~~~~&~~~~ $\Delta M_i [3293]$~~ &~~ $\Gamma_i [3240]$ ~~~~&~~~~ $\Delta M_i [3325]$ ~~& ~~ $\Gamma_i [3271]$ ~~~~&~~~~ $\Delta M_i [3332]$ ~~&~~ $\Gamma_i [3288]$ \\
\hline
$D K$ & -2.3 & 0.2 & $\cdots$ & $\cdots$ & $\cdots$ & $\cdots$ & -2.7 & 6.8 \\
$D_s \eta$ & -0.4 & $6.6 \times 10^{-3}$ & $\cdots$ & $\cdots$ & $\cdots$ & $\cdots$ & -0.3 & 0.7 \\
$D_s \eta'$ & -0.5 & $8.5 \times 10^{-5}$ & $\cdots$ & $\cdots$ & $\cdots$ & $\cdots$ & -0.4 & 0.2 \\
$D^* K$ & -1.2 & 0.2 & -1.3 & 0.5 & -4.0 & 2.6 & -3.8 & 5.0 \\
$D_s^* \eta$ & -0.2 & $1.0 \times 10^{-2}$ & -0.5 & 0.2 & -0.4 & 0.1 & -0.4 & 0.4 \\
$D_s^* \eta'$ & -0.1 & $1.4 \times 10^{-2}$ & -0.9 & 0.4 & -0.6 & $2.4 \times 10^{-2}$ & -0.5 & $1.3 \times 10^{-2}$ \\
$D_0(2550) K$ & -1.1 & $5.4 \times 10^{-3}$ & $\cdots$ & $\cdots$ & $\cdots$ & $\cdots$ & -2.4 & 4.1 \\
$D_{s0}(2590) \eta$ & -0.2 & $1.8 \times 10^{-2}$ & $\cdots$ & $\cdots$ & $\cdots$ & $\cdots$ & -0.3 & 0.1 \\
$D_1^*(2600) K$ & -0.9 & 0.3 & -4.6 & 1.0 & -8.7 & 4.0 & -4.2 & 2.9 \\
$D_{s1}^*(2700) \eta$ & $\cdots$ & $\cdots$ & -0.6 & 0.0 & -0.8 & $8.7 \times 10^{-4}$ & -0.3 & $9.0 \times 10^{-4}$ \\
$D_0^*(2300) K$ & $\cdots$ & $\cdots$ & -0.3 & 1.9 & -0.3 & $4.8 \times 10^{-2}$ & $\cdots$ & $\cdots$ \\
$D_{s0}^*(2317) \eta$ & $\cdots$ & $\cdots$ & $-3.5 \times 10^{-2}$ & 0.2 & -0.1 & 0.1 & $\cdots$ & $\cdots$ \\
$D_{s0}^*(2317) \eta'$ & $\cdots$ & $\cdots$ & $-3.5 \times 10^{-2}$ & 0.0 & -0.2 & 0.0 & $\cdots$ & $\cdots$ \\
$D_1(2430) K$ & -1.2 & 3.0 & -0.6 & 1.9 & 0.1 & 0.3 & -0.5 & $2.0 \times 10^{-2}$ \\
$D_{s1}(2460) \eta$ & -0.1 & 0.1 & -0.1 & 0.1 & $-2.6 \times 10^{-2}$ & 0.1 & -0.1 & $4.9 \times 10^{-4}$ \\
$D_1(2420) K$ & -44.9 & 64.0 & -0.5 & $2.2 \times 10^{-4}$ & -2.0 & 2.5 & -2.4 & 2.2 \\
$D_{s1}(2536) \eta$ & -5.5 & 2.1 & -0.1 & $2.6 \times 10^{-3}$ & -0.4 & 0.2 & -0.2 & $3.4 \times 10^{-2}$ \\
$D_2^*(2460) K$ & -2.1 & 3.8 & -19.1 & 44.9 & -6.2 & 7.6 & -3.9 & 3.2 \\
$D_{s2}^*(2573) \eta$ & -0.2 & $2.7 \times 10^{-2}$ & -3.3 & 3.2 & -0.6 & 0.3 & -0.5 & 0.2 \\
$D(1^3D_1)(2648) K$ & -7.1 & 1.3 & -1.1 & $4.9 \times 10^{-2}$ & -18.3 & 25.4 & -0.2 & 0.1 \\
$D(1D_2)(2723) K$ & $\cdots$ & $\cdots$ & $\cdots$ & $\cdots$ & -1.3 & 0.0 & -11.4 & 0.0 \\
$D_3^*(2750) K$ & $\cdots$ & $\cdots$ & -18.2 & 0.0 & -7.3 & $3.6 \times 10^{-2}$ & -8.3 & 3.3 \\
$D K^*$ & $-3.3 \times 10^{-2}$ & $5.9 \times 10^{-3}$ & -0.4 & 2.8 & 0.1 & 0.4 & -0.1 & 0.4 \\
$D_s \phi$ & $-6.1 \times 10^{-3}$ & $2.6 \times 10^{-6}$ & -0.2 & 0.6 & $-1.4 \times 10^{-2}$ & 0.2 & $-3.4 \times 10^{-2}$ & $2.8 \times 10^{-2}$ \\
$D^* K^*$ & -0.7 & 0.1 & -0.8 & 1.1 & -0.4 & 0.8 & -0.3 & 1.4 \\
$D_s^* \phi$ & -0.1 & $8.5 \times 10^{-3}$ & -0.2 & 0.1 & -0.1 & 0.1 & -0.1 & 0.3 \\
$D_0^*(2300) K^*$ & $\cdots$ & $\cdots$ & $-2.9 \times 10^{-2}$ & $1.8 \times 10^{-6}$ & -0.4 & $5.7 \times 10^{-3}$ & $-1.5 \times 10^{-2}$ & $3.2 \times 10^{-3}$ \\
$D_1(2430) K^*$ & $\cdots$ & $\cdots$ & $\cdots$ & $\cdots$ & -0.3 & 0.0 & -0.6 & 0.0 \\
$D_1(2420) K^*$ & $\cdots$ & $\cdots$ & $\cdots$ & $\cdots$ & -1.7 & 0.0 & -0.2 & 0.0 \\
\hline
$\mathrm{Total}$ & -68.8 & 75.2 & -52.9 & 59.0 & -53.9 & 44.8 & -44.1 & 31.4 \\
\hline\hline
\end{tabular}
\end{table*}

\begin{table*}[t]
\caption{Coupled-channel contributions to the mass shifts $\Delta M_i$ and OZI-allowed partial widths $\Gamma_i$ for the $D(1F)$ states. The notation and units are the same as in Table~\ref{tab:D_3S_channel_ccef}.}
\label{tab:D_1F_channel_ccef}
\small
\setlength{\tabcolsep}{1.15pt}
\renewcommand{\arraystretch}{1.05}
\begin{tabular}{cccccccccccccccccccccccccccccccccccccc}
\hline\hline
 & \multicolumn{2}{c}{\rule[-0.5ex]{0pt}{2.6ex}$\underline{~~~~~~~~~~~~~~D(1^3F_2)~~~~~~~~~~~~~~}$} & \multicolumn{2}{c}{\rule[-0.5ex]{0pt}{2.6ex}$\underline{~~~~~~~~~~~~~~D(1F_3)~~~~~~~~~~~~~~}$} & \multicolumn{2}{c}{\rule[-0.5ex]{0pt}{2.6ex}$\underline{~~~~~~~~~~~~~~D(1F_3')~~~~~~~~~~~~~~}$} & \multicolumn{2}{c}{\rule[-0.5ex]{0pt}{2.6ex}$\underline{~~~~~~~~~~~~~~D(1^3F_4)~~~~~~~~~~~~~~}$} \\
%\cline{2-3}\cline{4-5}\cline{6-7}\cline{8-9}
Channel &~~ $\Delta M_i [3125]$ ~~&~~ $\Gamma_i [2990]$ ~~~~&~~~~ $\Delta M_i [3078]$ ~~&~~ $\Gamma_i [2965]$ ~~~~&~~~~ $\Delta M_i [3167]$ ~~&~~ $\Gamma_i [3079]$ ~~~~&~~~~ $\Delta M_i [3092]$ ~~&~~ $\Gamma_i [3014]$ \\
\hline
$D \pi$ & -0.2 & 3.8 & $\cdots$ & $\cdots$ & $\cdots$ & $\cdots$ & -1.6 & 4.2 \\
$D_s K$ & -0.1 & 0.4 & $\cdots$ & $\cdots$ & $\cdots$ & $\cdots$ & -0.4 & 0.5 \\
$D \eta$ & $-3.5 \times 10^{-2}$ & 0.2 & $\cdots$ & $\cdots$ & $\cdots$ & $\cdots$ & -0.2 & 0.4 \\
$D \eta'$ & $-2.5 \times 10^{-2}$ & $4.1 \times 10^{-4}$ & $\cdots$ & $\cdots$ & $\cdots$ & $\cdots$ & -0.1 & $2.1 \times 10^{-2}$ \\
$D^* \pi$ & -0.3 & 1.3 & -0.7 & 2.3 & -6.5 & 14.4 & -3.4 & 10.7 \\
$D_s^* K$ & -0.1 & $4.5 \times 10^{-2}$ & -0.3 & 0.1 & -1.6 & 1.2 & -0.7 & 0.4 \\
$D^* \eta$ & $-5.0 \times 10^{-2}$ & $3.0 \times 10^{-2}$ & -0.1 & $4.4 \times 10^{-2}$ & -1.0 & 1.2 & -0.5 & 0.4 \\
$D^* \eta'$ & $-2.5 \times 10^{-2}$ & $5.4 \times 10^{-4}$ & -0.1 & $1.7 \times 10^{-8}$ & -0.5 & $1.4 \times 10^{-2}$ & -0.2 & $2.2 \times 10^{-4}$ \\
$D_0(2550) \pi$ & -1.8 & 1.2 & $\cdots$ & $\cdots$ & $\cdots$ & $\cdots$ & -3.5 & 0.2 \\
$D_{s0}(2590) K$ & -0.4 & 0.0 & $\cdots$ & $\cdots$ & $\cdots$ & $\cdots$ & -1.0 & 0.0 \\
$D_0(2550) \eta$ & -0.2 & 0.0 & $\cdots$ & $\cdots$ & $\cdots$ & $\cdots$ & $\cdots$ & $\cdots$ \\
$D_1^*(2600) \pi$ & -1.3 & 0.5 & -2.6 & 0.6 & -7.2 & 0.4 & -3.6 & 0.1 \\
$D_0^*(2300) \pi$ & $\cdots$ & $\cdots$ & -0.1 & $4.5 \times 10^{-3}$ & -3.8 & 7.9 & $\cdots$ & $\cdots$ \\
$D_{s0}^*(2317) K$ & $\cdots$ & $\cdots$ & $-2.3 \times 10^{-2}$ & $8.6 \times 10^{-4}$ & -1.3 & 1.0 & $\cdots$ & $\cdots$ \\
$D_0^*(2300) \eta$ & $\cdots$ & $\cdots$ & $-1.4 \times 10^{-2}$ & $8.9 \times 10^{-5}$ & -0.5 & 0.2 & $\cdots$ & $\cdots$ \\
$D_1(2430) \pi$ & -1.0 & $4.2 \times 10^{-2}$ & -0.7 & 0.1 & -3.0 & 5.8 & -6.9 & 8.8 \\
$D_{s1}(2460) K$ & -0.2 & $4.3 \times 10^{-4}$ & -0.2 & $9.0 \times 10^{-7}$ & -0.9 & 0.1 & -1.8 & $2.3 \times 10^{-2}$ \\
$D_1(2430) \eta$ & -0.1 & $2.6 \times 10^{-4}$ & -0.1 & $2.9 \times 10^{-8}$ & -0.4 & 0.1 & -0.7 & $1.1 \times 10^{-2}$ \\
$D_1(2420) \pi$ & -17.5 & 1.8 & -0.7 & 0.1 & -8.7 & 7.2 & -5.5 & 1.1 \\
$D_{s1}(2536) K$ & -3.8 & 0.0 & -0.1 & 0.0 & -1.9 & $5.7 \times 10^{-3}$ & -1.2 & 0.0 \\
$D_1(2420) \eta$ & -1.9 & 0.1 & -0.1 & 0.0 & -1.0 & 0.1 & -0.6 & $6.3 \times 10^{-4}$ \\
$D_2^*(2460) \pi$ & -4.4 & 0.9 & -16.7 & 4.7 & -11.7 & 3.4 & -11.7 & 4.6 \\
$D_{s2}^*(2573) K$ & -1.0 & 0.0 & -3.6 & 0.0 & -2.6 & $6.3 \times 10^{-3}$ & -2.7 & 0.0 \\
$D_2^*(2460) \eta$ & -0.4 & 0.0 & -1.6 & 0.0 & -1.4 & $4.0 \times 10^{-2}$ & -1.3 & $2.5 \times 10^{-6}$ \\
$D(2^3P_0)(2673) \pi$ & $\cdots$ & $\cdots$ & -1.8 & $5.2 \times 10^{-3}$ & -1.9 & 0.8 & $\cdots$ & $\cdots$ \\
$D(2P_1)(2775) \pi$ & -9.3 & 1.9 & -1.4 & $1.3 \times 10^{-5}$ & -0.9 & $1.5 \times 10^{-3}$ & -3.2 & $2.3 \times 10^{-2}$ \\
$D(2P_1^\prime)(2830) \pi$ & -11.4 & 0.5 & -1.3 & 0.0 & -2.5 & $3.0 \times 10^{-2}$ & -2.3 & $1.0 \times 10^{-4}$ \\
$D(2^3P_2)(2901) \pi$ & -2.6 & 0.0 & -13.3 & 0.0 & -3.9 & $3.2 \times 10^{-2}$ & -2.5 & 0.0 \\
$D(1^3D_1)(2648) \pi$ & -1.1 & 0.1 & -1.1 & 0.1 & -3.2 & 5.9 & -0.2 & $3.2 \times 10^{-3}$ \\
$D(1D_2)(2723) \pi$ & -3.9 & 1.5 & -3.4 & $1.6 \times 10^{-4}$ & -1.4 & 0.1 & -3.7 & 0.2 \\
$D(1D'_2)(2789) \pi$ & -63.9 & 0.0 & -1.4 & 0.0 & -10.1 & 0.1 & -5.2 & $2.4 \times 10^{-5}$ \\
$D_3^*(2750) \pi$ & -5.3 & $2.8 \times 10^{-2}$ & -60.8 & 49.6 & -9.4 & 3.0 & -12.6 & 0.2 \\
$D \rho$ & $-3.5 \times 10^{-2}$ & $3.9 \times 10^{-2}$ & -0.2 & 0.7 & $8.6 \times 10^{-3}$ & 0.1 & $-3.7 \times 10^{-2}$ & 0.1 \\
$D \omega$ & $-1.1 \times 10^{-2}$ & $1.2 \times 10^{-2}$ & -0.1 & 0.2 & $2.4 \times 10^{-3}$ & $4.7 \times 10^{-2}$ & $-1.2 \times 10^{-2}$ & $1.8 \times 10^{-2}$ \\
$D_s K^*$ & $-6.8 \times 10^{-3}$ & $3.2 \times 10^{-4}$ & -0.1 & $3.2 \times 10^{-2}$ & $-1.3 \times 10^{-2}$ & 0.1 & $-7.4 \times 10^{-3}$ & $8.1 \times 10^{-4}$ \\
$D^* \rho$ & -0.9 & 1.1 & -0.4 & 0.4 & -0.2 & 0.6 & -0.2 & 0.5 \\
$D^* \omega$ & -0.3 & 0.3 & -0.1 & 0.1 & -0.1 & 0.2 & -0.1 & 0.2 \\
$D_s^* K^*$ & -0.2 & 0.0 & -0.1 & 0.0 & -0.1 & $2.3 \times 10^{-2}$ & $-4.8 \times 10^{-2}$ & $3.2 \times 10^{-4}$ \\
$D_0^*(2300) \rho$ & -1.4 & 0.0 & $\cdots$ & $\cdots$ & -0.1 & 0.0 & $\cdots$ & $\cdots$ \\
$D_0^*(2300) \omega$ & $\cdots$ & $\cdots$ & $\cdots$ & $\cdots$ & $-3.7 \times 10^{-2}$ & 0.0 & $\cdots$ & $\cdots$ \\
\hline
$\mathrm{Total}$ & -135.2 & 15.8 & -113.2 & 59.1 & -87.8 & 54.1 & -77.7 & 32.7 \\
\hline\hline
\end{tabular}
\end{table*}

\begin{table*}[t]
\caption{Coupled-channel contributions to the mass shifts $\Delta M_i$ and OZI-allowed partial widths $\Gamma_i$ for the $D_s(1F)$ states. The notation and units are the same as in Table~\ref{tab:D_3S_channel_ccef}.}
\label{tab:Ds_1F_channel_ccef}
\small
\setlength{\tabcolsep}{1.25pt}
\renewcommand{\arraystretch}{1.05}
\begin{tabular}{cccccccccccccccccccccccccc}
\hline\hline
 & \multicolumn{2}{c}{\rule[-0.5ex]{0pt}{2.6ex}$\underline{~~~~~~~~~~~~~~D_s(1^3F_2)~~~~~~~~~~~~~~}$} & \multicolumn{2}{c}{\rule[-0.5ex]{0pt}{2.6ex}$\underline{~~~~~~~~~~~~~~D_s(1F_3)~~~~~~~~~~~~~~}$} & \multicolumn{2}{c}{\rule[-0.5ex]{0pt}{2.6ex}$\underline{~~~~~~~~~~~~~~D_s(1F_3')~~~~~~~~~~~~~~}$} & \multicolumn{2}{c}{\rule[-0.5ex]{0pt}{2.6ex}$\underline{~~~~~~~~~~~~~~D_s(1^3F_4)~~~~~~~~~~~~~~}$} \\
%\cline{2-3}\cline{4-5}\cline{6-7}\cline{8-9}
Channel &~~ $\Delta M_i [3182]$ ~~&~~ $\Gamma_i [3154]$~~~~ &~~~~ $\Delta M_i [3152]$ ~~&~~ $\Gamma_i [3128]$ ~~~~&~~~~ $\Delta M_i [3224]$ ~~&~~ $\Gamma_i [3176]$ ~~~~&~~~~ $\Delta M_i [3166]$ ~~& ~~$\Gamma_i [3125]$ \\
\hline
$D K$ & $-4.8 \times 10^{-2}$ & 2.3 & $\cdots$ & $\cdots$ & $\cdots$ & $\cdots$ & -1.6 & 4.6 \\
$D_s \eta$ & $-2.8 \times 10^{-2}$ & 0.1 & $\cdots$ & $\cdots$ & $\cdots$ & $\cdots$ & -0.2 & 0.3 \\
$D_s \eta'$ & -0.1 & $9.8 \times 10^{-3}$ & $\cdots$ & $\cdots$ & $\cdots$ & $\cdots$ & -0.2 & $3.5 \times 10^{-2}$ \\
$D^* K$ & -0.1 & 0.6 & -0.2 & 1.3 & -6.4 & 14.3 & -3.4 & 6.4 \\
$D_s^* \eta$ & $-3.8 \times 10^{-2}$ & $9.5 \times 10^{-3}$ & -0.1 & $2.0 \times 10^{-2}$ & -0.7 & 0.8 & -0.3 & 0.3 \\
$D_s^* \eta'$ & -0.1 & $1.5 \times 10^{-2}$ & -0.1 & $2.0 \times 10^{-2}$ & -0.7 & $1.3 \times 10^{-2}$ & -0.3 & $4.3 \times 10^{-4}$ \\
$D_0(2550) K$ & -1.5 & 0.7 & $\cdots$ & $\cdots$ & $\cdots$ & $\cdots$ & -3.2 & $8.1 \times 10^{-3}$ \\
$D_{s0}(2590) \eta$ & -0.2 & $1.1 \times 10^{-3}$ & $\cdots$ & $\cdots$ & $\cdots$ & $\cdots$ & -0.4 & 0.0 \\
$D_1^*(2600) K$ & -1.1 & $3.3 \times 10^{-2}$ & -2.4 & $1.9 \times 10^{-3}$ & -6.2 & $8.4 \times 10^{-3}$ & -3.3 & $2.0 \times 10^{-8}$ \\
$D_0^*(2300) K$ & $\cdots$ & $\cdots$ & -0.1 & $1.2 \times 10^{-2}$ & -4.4 & 6.6 & $\cdots$ & $\cdots$ \\
$D_{s0}^*(2317) \eta$ & $\cdots$ & $\cdots$ & $-1.3 \times 10^{-2}$ & $1.8 \times 10^{-3}$ & -0.6 & 0.8 & $\cdots$ & $\cdots$ \\
$D_1(2430) K$ & -1.4 & 0.1 & -1.0 & 0.1 & -3.4 & 4.0 & -7.6 & 5.4 \\
$D_{s1}(2460) \eta$ & -0.1 & $1.6 \times 10^{-3}$ & -0.1 & $2.5 \times 10^{-3}$ & -0.4 & 0.2 & -1.0 & 0.1 \\
$D_1(2420) K$ & -13.6 & 13.1 & -0.6 & 0.1 & -8.8 & 5.1 & -5.1 & 0.6 \\
$D_{s1}(2536) \eta$ & -2.0 & 0.8 & -0.1 & $8.5 \times 10^{-5}$ & -0.9 & $2.8 \times 10^{-2}$ & -0.5 & $4.6 \times 10^{-4}$ \\
$D_2^*(2460) K$ & -3.8 & 2.5 & -14.9 & 17.7 & -10.6 & 2.1 & -11.5 & 2.1 \\
$D_{s2}^*(2573) \eta$ & -0.5 & 0.1 & -1.9 & 0.1 & -1.1 & $2.9 \times 10^{-2}$ & -1.3 & $1.1 \times 10^{-5}$ \\
$D(1^3D_1)(2648) K$ & -2.1 & $2.0 \times 10^{-6}$ & -1.3 & 0.0 & -3.8 & 0.2 & -0.2 & 0.0 \\
$D K^*$ & -0.1 & 0.1 & -0.2 & 1.1 & $5.5 \times 10^{-3}$ & 0.2 & -0.1 & 0.1 \\
$D_s \phi$ & $-8.4 \times 10^{-3}$ & $1.2 \times 10^{-3}$ & -0.1 & 0.1 & $-1.3 \times 10^{-2}$ & $4.7 \times 10^{-2}$ & $-7.9 \times 10^{-3}$ & $7.2 \times 10^{-4}$ \\
$D^* K^*$ & -1.2 & 2.3 & -0.6 & 0.9 & -0.3 & 0.8 & -0.2 & 0.7 \\
$D_s^* \phi$ & -0.2 & $3.2 \times 10^{-3}$ & -0.1 & 0.0 & -0.1 & $7.5 \times 10^{-3}$ & $-4.5 \times 10^{-2}$ & 0.0 \\
\hline
$\mathrm{Total}$ & -28.2 & 22.8 & -23.8 & 21.5 & -48.4 & 35.2 & -40.5 & 20.6 \\
\hline\hline
\end{tabular}
\end{table*}

\bibliography{ref}

\end{document}